\documentclass[12pt, draftclsnofoot, onecolumn]{IEEEtran}
\usepackage{blindtext}
\usepackage{graphicx,hyperref}
\usepackage{bigints}
\usepackage{caption,float}
\newlength\figureheight 
\newlength\figurewidth 
\usepackage{cite}
\usepackage{url}
\usepackage{blindtext, graphicx}
\usepackage{amsthm,amsmath,amssymb,mathtools,bm,etoolbox}

\usepackage{cuted}
\usepackage{multicol,lipsum}
\usepackage{multirow}
\usepackage{breqn}
\newcommand\ddfrac[2]{\frac{\displaystyle #1}{\displaystyle #2}}
\usepackage[dvipsnames]{xcolor}
\newtheorem{rem}{Remark}
\usepackage{boxhandler}
\usepackage{subcaption,lipsum}
\newtheorem{thm}{Theorem}
\newcommand{\e}[1]{{\mathbb E}\left[ #1 \right]}
\usepackage{stackengine}
\def\delequal{\mathrel{\ensurestackMath{\stackon[1pt]{=}{\scriptstyle\Delta}}}}
\usepackage{suffix}

\DeclarePairedDelimiterX\MeijerM[3]{\lparen}{\rparen}%
{\begin{smallmatrix}#1 \\ #2\end{smallmatrix}\delimsize\vert\,#3}

\newcommand\MeijerG[8][]{%
  G^{\,#2,#3}_{#4,#5}\MeijerM[#1]{#6}{#7}{#8}}

\WithSuffix\newcommand\MeijerG*[7]{%
  G^{\,#1,#2}_{#3,#4}\MeijerM*{#5}{#6}{#7}}
\usepackage{ellipsis}
\usepackage{tkz-euclide}
\usepackage{tikz}
\usepackage{tikz-3dplot}
\usepackage[utf8]{inputenc}
\usepackage{pgfplots} 
\usepackage{pgfgantt}
\usepackage{pdflscape}
\pgfplotsset{compat=newest} 
\pgfplotsset{plot coordinates/math parser=false}
\pgfplotsset{every  tick/.style={black,},ylabel style={font=\tiny},xlabel style={font=\tiny},tick label style={font=\tiny},legend style= {font=\scriptsize},
minor x tick num=1,minor y tick num=1,xminorticks=true,yminorticks=true,}
  \newlength\fheight
\newlength\fwidth

\begin{document}
\title{\textcolor{black}{Tractable Approach to MmWaves Cellular Analysis with FSO Backhauling under Feedback Delay and Hardware Limitations}}
\author{Elyes~Balti,~\IEEEmembership{Student Member,~IEEE,}
        and~Brian~K.~Johnson,~\IEEEmembership{Fellow,~IEEE}}%
\maketitle

\begin{abstract}
\textcolor{black}{In this work, we investigate the performance of a millimeter waves (mmWaves) cellular system with free space optical (FSO) backhauling. MmWave channels are subject to Nakagami-m fading while the optical links experience the Double Generalized Gamma including atmospheric turbulence, path loss and the misalignment  between the transmitter and the receiver aperture (also known as the pointing errors). The FSO model also takes into account the receiver detection technique which could be either heterodyne or intensity modulation and direct detection (IM/DD). Each user equipment (UE) has to be associated to one serving base station (BS) based on the received signal strength (RSS) or Channel State Information (CSI). We assume partial relay selection (PRS) with CSI based on mmWaves channels to select the BS associated with the highest received CSI. Each serving BS decodes the received signal for denoising, converts it into modulated FSO signal, and then forwards it to the data center. Thereby, each BS can be viewed as a  decode-and-forward (DF) relay. In practice, the relay hardware suffers from nonlinear high power amplification (HPA) impairments which, substantially degrade the system performance. In this work, we will discuss the impacts of three common HPA impairments named respectively, soft envelope limiter (SEL), traveling wave tube amplifier (TWTA), and solid state power amplifier (SSPA). Novel closed-forms and tight upper bounds of the outage probability, the probability of error, and the achievable rate are derived}. Capitalizing on these performance, we derive the high SNR asymptotes to get engineering insights into the system gain such as the diversity order.
\end{abstract}

\IEEEpeerreviewmaketitle

\section{Introduction}
\IEEEPARstart{W}{ith} the rapid increase of the internet base, the mobile stations, and the extremely high demand for bandwidth, the microwave Radio Frequency (RF) cellular systems have reached a saturation level owing to the limited spectrum, and expensive access licence. Although, many research attempts in cognitive radio allow parallel utilization of the bandwidth between the primary and secondary users, the last ones still suffer from the spectrum drought since they are always leveraging from some spectrum holes left by the primary users. Moreover, the backhaul network cannot support the big data flow even for the licenced primary ones.

Moreover, the backhaul network cannot support the big data flow even for the licenced primary ones. Recent attempts have proposed the usage of optical fibers (OF) as a solution for the backhaul network congestion. However, for ultra dense cellular networks, a large number of OF are needed given that these cable installations are very costly and the space installation of such cables to serve large number of cells/users are limited and even restricted in some areas.
\subsection{Motivation}
\textcolor{black}{To support such network densification, millimeter waves (mmWaves) technology, which refers to the spectrum from 28 to 300 GHz, has emerged as a promising solution to replace microwave communications. In fact, mmWaves provide a large available spectrum and increase the cellular network capacity thanks to the high bandwidth offered by such technology \cite{surv}. In addition, mmWaves technology is becoming practical and has available commercial products such as IEEE 802.11 ad wireless gigabit alliance (WiGig), 5G modem, and 5G NR mmwave prototype, etc \cite{rheath,5gnr}. In addition, mmWaves technology has tremendous applications in the vehicular communications area, in particular, for self-driving vehicles requiring a big data exhange with road side units (RSU) to enhance the vehicle awareness and avoid potential accidents \cite{v1,v2}}.
 
Free Space Optical (FSO) communications has been recently proposed as an alternative or complementary solution to both RF and OF due to its flexibility, free spectrum access licence, power efficiency, cost effectiveness, no installation restriction and most importantly it is a way to densify the cellular networks with limited congestion and delays \cite{5,8}. Due to these advantages, FSO is seen as the corner stone of the-fifth generation since it is predicted to achieve 25 times the average cell throughput, 10 times the spectral/energy efficiency, 1000 times the system capacity and from 10 to 100 times the data rate compared to the LTE or the fourth-generation (4G) \cite{10,mmwave}. Besides, FSO systems employ a narrow laser beam which offers a high security level, immunity to electromagnetic interference and operating frequencies above 300 GHz. Because of these advantages, FSO technology has been considered as a possible solution for the last mile problem to bridge the bandwidth gap between the end-users and the OF backbone network. Moreover, the FSO technology has been also applied over the following applications such as enterprise/campus connectivity, video surveillance and monitoring, redundant link and disaster recovery, security and broadcasting \cite{5}.

To improve the coverage and the scalability of the network, one solution is to implement the relays between the transmitter (Tx) and the receiver (Rx). Because of this advantage, cooperative relaying-assisted communication is considered as one of the key technologies for the next generation wireless communications because it plays an important role in improving the Quality of Service (QoS), reliability and coverage \cite{11}. The majority of the research attempts investigated mixed RF/FSO system considering various relaying schemes. The most prominent ones are Amplify-and-Forward (AF) \cite{aggregate,glob2017,61}, Decode-and-Forward (DF) \cite{17}, Quantize-and-Encode (QE) \cite{19}, and Quantize-and-Forward (QF) \cite{20}. Moreover, many research attempts have assumed systems employing either single or multiple relays. For the single relay system, there is only one way to forward the signal to the destination through the relay. \textcolor{black}{For serial multiple relays, also called multihop relaying systems, recent works have investigated this topic. Specifically, the works \cite{hop1,hop2,hop3} have studied the performance analysis of multihop relaying systems, in particular, they derived the probability of outage, and the ergodic capacity. In addition, the work \cite{hop3} also derived the performance analysis of mesh networks and compared the outcomes with those of mixed RF/FSO multihop relaying systems while assuming nonlinear relay power amplifiers}. \textcolor{black}{For parallel deployment of multiple relays}, there are two possible ways either sending parallel transmissions when simultaneously activate all the relays or selecting one relay among the total set. In fact, there are many relay selection protocols such as opportunistic relay selection, partial relay selection \cite{21}, distributed switch and stay, max-select protocol and all active relaying \cite{24}. \textcolor{black}{The latter is not convenient since the optical front-end receiver will experience synchronization problems.}
\subsection{Literature}
\textcolor{black}{For reliable communications, mmWaves technology is dedicated for short range communications. Due to its high frequencies, mmWaves suffers from the severe pathloss experienced during the free space propagation. The link budgets become more subject to degradation when the distance between the Tx and the Rx gets larger since the received power is inversely proportional to the distance. To address this shortcoming, mmWaves systems involve the implementation of multiple antennas to provide an additional array gain in order to compensate for the pathloss severity. Also the multiple antennas setting reduces the effects of the interference by using high directional antennas or sectorized arrays. Moreover, mmWaves systems usually achieve low signal-to-interference-plus-noise ratio (SINR) as the bandwidth is very high, yielding a severe noise power at the Rx. On the other side, mmWaves systems are mainly characterized by the high achievable rate which is the main motivation behind the introduction of the fifth generation (5G). For this purpose, mmWaves links are employed in dense microcells where low power and high data rate in Gbps are required for exchanging flow data between the UEs, whereas sub-6 GHz is used for macrocells where low data rate and high power links are required to exchange the data between the long-distant macrocells.}

Previous work have proposed various channel models for the optical fading. In fact, Lognormal distribution is widely employed to statistically model the optical irradiance \cite{ln1} since it provides a good fit to the experimental data for weak turbulence. However, Lognormal model largely deviates from the experimental data as the atmospheric turbulence becomes more severe. To overcome this shortcoming, recent work have proposed the so-called Gamma-Gamma (G$^2$) \cite{60} as a model for the FSO fading since it provides a good fit to the experimental data for a wider range of the atmospheric turbulences compared to the Log-normal distribution. However, G$^2$ fails to provide a good fit with the experimental data especially at the tails. Since the calculation of the fade and the detection probability are essentially based on the tail of the probability density function (pdf), underestimation or overestimation of the tail region affects the performance analysis accuracy and certainly leads to erroneous results. To address this problem, Kashani \textit{et. al} \cite{dgg} introduced a new efficient optical fading model called Double Generalized Gamma (DGG) which not only reflects a wide range of the atmospheric turbulences but also it provides a good fit to the experimental data particularly at the tail region.

As the optical signal propagates in free-space, it is susceptible not only to the atmospheric turbulences but also to the path loss and the pointing errors as well. The path loss is basically depends on the link distance and the atmospheric attenuation which describes the weather conditions going from clear air, hazy, rainy and foggy. The work \cite{ln1,43} provide some typical values of the atmospheric attenuations describing the corresponding weather conditions. Moreover, the optical signal is also subject to the pointing errors which can be described as the misalignment between the laser-emitting relay and the receiver photodetector. In fact, this misalignment is mainly caused by the building sway and seismic activities resulting in the pointing errors that may arise severely as the relays and the receiver are located on tall buildings. The pointing errors can be interpreted as an additional FSO fading that requires an accurate model to quantify its impact on the FSO signal. Uysal \textit{et. al} \cite{44} have proposed various models for the radial displacement of the pointing errors assuming a Gaussian laser beam. The most general model proposed is called Beckmann pointing errors model and there are various special cases derived from it. Previous work have assumed that the radial displacement can be modeled as Rician \cite{45}, Hoyt \cite{46}, Non-Zero-Mean and Zero-Mean Single-Sided Gaussian \cite{47} but the most prevalent one is Rayleigh \cite{48,int} for simplicity. 


Regarding the HPA non linearities, this impairment is originated by the non linear relaying amplification resulting in a non linear distortion is created and affects substantially the quality of the signal. In practice, there is a finite maximum output level for which any power amplifier can produce it and such saturation level is basically amplifier-dependent and varies to some extent but regardless of the amplifier model, this ceiling level is always bounded. In case when the power amplifier becomes unable to produce such power level, a signal distortion over the peak may arise and such phenomena is called clipping (clipping factor) of the power amplifier. In addition, the HPA model can be classified into two categories which are memoryless HPA and HPA with memory. The HPA is considered memoryless or frequency-independent if its frequency response characteristics are flat over the operating frequency range and in this case, the HPA is fully characterized by the two characteristics AM/AM (amplitude to amplitude conversion) and AM/PM (amplitude to phase conversion). On the other hand, the HPA is said to be with memory if its frequency responce characteristics are totally dependent on the frequency components or to the thermal phenomena \cite{42}. Such model can be classified as Hammerstein system that can be modeled by a series of a memoryless HPA and a linear filter. There are many types of this impairment that have been already covered in the literature but the most widely used are Soft Envelope Limiter (SEL), Traveling Wave Tube Amplifier (TWTA) and Solid State Power Amplifier (SSPA) or also called the Rapp model \cite{28}. The SEL is typically used to model a HPA with a perfect predistortion system while the TWTA has been primarily considered to model the non linearities effect in OFDM system. However, the SSPA is characterized by a smoothness factor to control the switching between the saturation and the linear ranges. This model effectively discusses a linear characteristic for low magnitudes of the input signal and then it is limited by a definite constant saturated output. As the smoothness factor grows largely to infinity, this HPA model becomes the SEL impairments model. 
\subsection{Contribution}
\textcolor{black}{In this paper, we provide a global framework analysis where the communication between the UE and the data center takes two time slots. The first slot corresponds to uplink mmWaves cellular communications between the UE and its serving BS. In the second slot, the BS/relay forwards the signal after optical conversion and amplification to the front-end detector of the data center. To improve the coverage, after the re-encoding phase the BS/relay assists the signal by a high amplification gain which creates the signal distortion and originates the hardware impairments. In the same context, we study the effects of three HPA hardware impairments models which are the SEL, TWTA, and SSPA on the probability of outage, the probability of error, and the achievable rate. For each UE to select its serving BS, it receives periodic feedback from the nearest BSs. Under the assumptions of narrowband, fast fading channels, and slow propagation of the feedback, the UE will select its serving BS based on the outdated CSI rather than the update one. To model this delay based selection, we introduce the PRS protocol so that each UE can be associated to its serving BS. Moroever, we assume that each BS, interpreted as a decode-and-forward (DF) relay, collects the received signal following the maximal ratio combining (MRC). We also assume that the reference communication is altered by the interference coming from the other UEs.}
\textcolor{black}{To the best of our knowledge, this is the first work presenting a global framework analysis of mmWaves cellular systems with FSO backhauling under the hardware impairments, the interference, the blockage, and the feedback delay constraints. The analysis of this paper follows these steps:
\begin{itemize}
    \item Present a detailed description of the cellular network and the analysis of the outdated CSI, the interference, the blockage, and the hardware impairments.
    \item Define the channels models for mmWaves and FSO links along with the necessary parameters. Then derive the statistical distribution of all the channels.
    \item Analyze the signal-to-interference-plus-noise-plus-distortion ratio (SINDR).
    \item Derive the system performance metrics: the SINDR outage, the probability of error, the ergodic achievable rate, and the rate coverage.
    \item Derive the high signal-to-noise ratio (SNR) analytical asymptotes and characterize the diversity gain achieved by the system.
    \item Derive quantitative summaries and valuable engineering insights to draw meaningful conclusions and observations of the proposed system.
\end{itemize}
}
\subsection{Structure}
\textcolor{black}{This paper is organized as follows: Section II describes the cellular system model. The FSO backhauling analysis is presented in Section III while the performance analysis of the hybrid system is detailed in Section IV. Section V illustrates the numerical results with the discussion while the concluding remarks are summarized in Section VI.}
\subsection{Notation}
For the sake of organization, we provide some useful notations to avoid the repetition. $F_{h}(\cdot)$ and $f_{h}(\cdot)$ denote the cdf and pdf of the random variable $h$, respectively. The Generalized Gamma distribution with parameters $\alpha, \beta$ and $\gamma$ is denoted by $\mathcal{GG} (\alpha,\beta,\gamma)$ while the Gamma distribution of scale $\alpha$ and shape $\beta$ is denoted by $\mathcal{G}(\alpha,\beta)$. In addition, the Gaussian distribution of mean $\mu$ and variance $\sigma^2$ is denoted by $\mathcal{N}(\mu,\sigma^2)$. The operator $\e{\cdot}$ stands for the expectation while $\mathbb{P}$[$\cdot$] denotes the probability measure. The symbol $\backsim$ stands for "distributed as".
\section{MmWaves Cellular Network Analysis}
\begin{figure}[H]
\centering
\setlength\fheight{4cm}
\setlength\fwidth{9cm}
\begin{tikzpicture}

\draw [circle,circle,fill=gray!20!white] (.1,2) circle [radius=0.5] node {$\bold{BS_{1}}$};
\draw [circle,fill=gray!20!white](0.1,0) circle [radius=0.5] node {$\bold{BS_{k}}$};
\draw [circle,fill=gray!20!white](0.1,-2) circle [radius=0.5] node {$\bold{BS_{M}}$};
\draw [circle,circle,fill=gray!20!white](-4.2,0) circle [radius=0.5] node {$\bold{UE}$};
\draw [thick,dotted,blue] (0.1,1.3) -- (0.1,0.7);
\draw [thick,dotted,blue] (0.1,-0.7) -- (0.1,-1.3);

\draw[->,thick,blue] (-3.7,-0.1) -- node[above,sloped] {} node[below,sloped]{\small $\gamma_{1(k)}$} (-0.4,-0.1); 

\draw[<-,thick,dashed,blue] (-3.7,0.1) -- node[above,sloped] {\small $\hat{\gamma}_{1(k)}$} node[below,sloped]{} (-0.4,0.1);

\draw[->,thick,dotted,blue] (-3.8,0.32) -- node[above,sloped] {} node[below,sloped]{\small $\gamma_{1(1)}$} (-0.4,2); 

\draw[<-,thick,dashed,blue] (-4,0.4) -- node[above,sloped] {\small $\hat{\gamma}_{1(1)}$} node[below,sloped]{} (-0.4,2.2);

\draw[->,thick,dotted,blue] (-3.8,-0.45) -- node[above,sloped] {} node[below,sloped]{\small $\gamma_{1(M)}$} (-0.4,-2.2);

\draw[<-,thick,dashed,blue] (-3.8,-0.32) -- node[above,sloped] {\small $\hat{\gamma}_{1(M)}$} node[below,sloped]{} (-0.4,-2);




\draw[->,thick,blue] (1.5,0.5) -- (2,0.5);
\node[text width=5cm] at (4.7,0.5) {\small Active channel via selected relay}; 
\draw[->,dashed,thick,blue] (1.5,0) -- (2,0);
\node[text width=5cm] at (4.7,0) {\small CSI used for relay selection}; 
\draw[->,dotted,thick,blue] (1.5,-0.5) -- (2,-0.5);
\node[text width=5cm] at (4.7,-0.5) {\small Idle channels}; 
 
 
 
 
\end{tikzpicture}
    \caption{\textcolor{black}{Illustration of the PRS protocol. Among the set of the nearest BSs, the UE selects the BS associated with the strongest CSI. The average number of the nearest BSs to the reference UE is $M$.}}
    \label{prs}
\end{figure}
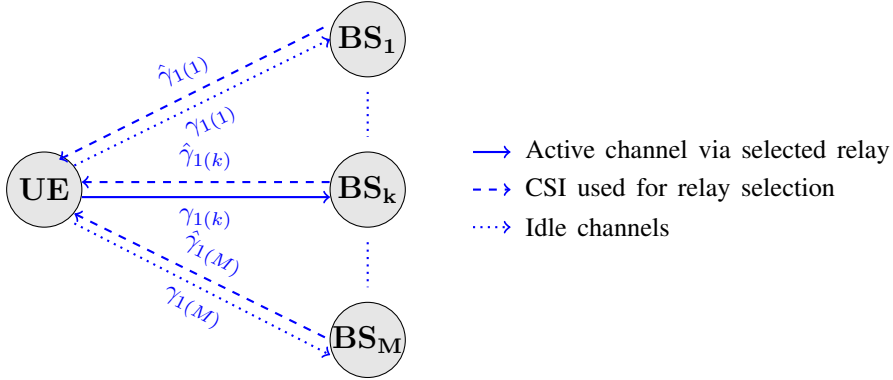

The system consists of Tx, Rx, and $N$ BSs/relays wirelessly linked to Tx (user) and Rx (data center) shown by Fig.~\ref{prs}. As mentioned earlier, these relays employ the DF relaying scheme to process the signal. To select the candidate relay of rank \textit{k}, we refer to the PRS with outdated CSI to pick the best one based on the local feedbacks of the RF channels.
\subsection{Relay Selection Protocol}
For a given communication, the UE receives local feedback ($\gamma_{1(i)}$ for $i = 1,\ldots M$) of the first hop obtained by the channel estimation from the \textit{M} relays and arranges them in an increasing order of amplitudes as follows: $\gamma_{1(1)}\leq\gamma_{1(2)}\leq \ldots \leq\gamma_{1(M)}$. The best scenario is to select the best relay \textit{(k = M)}. However, the best relay is not always available, so the UE will pick the next best available relay. Thus PRS consists of selecting the \textit{k}-th worst or \textit{(M - k)}-th best relay $R_{(k)}$. Given that the local feedback coming from the relays to the UE are very slow and the channels are time-varying, the CSI that is used for the relay selection is not the same as the CSI used for the transmission. In this case, an outdated CSI must be considered instead of the perfect one. As a result, the current and outdated CSI are correlated with the correlation coefficient $\rho$ as follows
\begin{equation}\label{corr}
\gamma_{1(k)} = \sqrt{\rho}~\hat{\gamma}_{1(k)} + \sqrt{1-\rho}~\omega_{k}, 
\end{equation}

where $k$ is the rank of the selected relay, $\omega_{k}$ is a random variable that follows the circularly complex Gaussian distribution with the same variance of the channel gain $h_{1(k)}$. The correlation coefficient $\rho$ is given by the Jakes' autocorrelation model \cite{jakes} as follows
\begin{equation}\label{eq:2}
\rho = J_0(2\pi f_{d} T_d), 
\end{equation}

where $J_0(\cdot)$ is the zeroth order Bessel function of the first kind, $T_d$ is the propagation delay of the feedback, and $f_d$ is the maximum Doppler frequency of the channels.

\subsection{System Model}
\textcolor{black}{In this scenario, we assume that the BSs are equipped with multiple antennas, $N$, while the UE has single antenna. With an appropriate Rx array gain to compensate the pathloss, the assumption of single antenna at the Tx seems to work while the results are confirmed by \cite{yuyang}. Given that MRC is assumed at the BS with equal power split for the Rx antennas, the received signal of the UE is expressed as
\begin{equation}
y = \sqrt{\frac{\Omega P}{N}}\sum_{n=1}^Nh_n s + \sum_{p=1}^{M_z}g_pd_p + \boldsymbol{n},    
\end{equation}}

\textcolor{black}{where $\Omega$ is the average path gain, $P$ is the transmit power of the UE, $s$ is the transmitted symbol, $h_n$ is the $n$-th channel gain, $g_p$ is the $p$-th interfering channel, $d_p$ is the transmitted symbol of the $p$-th interferer, $M_z$ is the average number of interferers, and $\boldsymbol{n}$ is the zero-mean additive white Gaussian noise (AWGN) with variance $\sigma_1^2$ given by
\begin{equation}
\sigma_1^2[\text{dBm}] = B + N_0 + N_f.    
\end{equation}}
\textcolor{black}{TABLE \ref{params} summarizes the values of the system parameters used in the simulation.} 
\begin{table}[H]
\renewcommand{\arraystretch}{1}
\caption{Cellular System Parameters}\label{params}
\label{table_example}
\centering
\begin{tabular}{|c|c|c|}
\hline
\bfseries Parameter & \bfseries Symbol  &  \bfseries Value \\
\hline
Carrier frequency & $f_c$   & 30 GHz\\
\hline
Transmit antenna element gain & $G_{t}$   & 4 dB\\
\hline
Receive antenna element gain & $G_{r}$   &  4 dB\\
\hline
Number of BS antennas & $N$   & 64\\
\hline
Noise spectral density & $N_0$   & -142 dBm/Hz\\
\hline
Noise figure & $N_f$   & 0 dB\\
\hline
Bandwidth & $B$   & 700 MHz\\
\hline
Speed of light & $c$   & 3 $10^8$ m/s\\
\hline
Link distance & $L_1$   & 50 m\\
\hline
\end{tabular}
\end{table}
\textcolor{black}{The path gain is given
\begin{equation}
\Omega[\text{dB}] = G_{t} + G_{r} -20\log_{10}\left(\frac{4\pi L_1 f_c}{c} \right) - \alpha L_1,
\end{equation}} 
\textcolor{black}{where $\alpha$ is the pathloss exponent}.
\subsection{Channel Model}
\textcolor{black}{Since all the channels between the UE and the BS antennas are  Nakagami-m distributed, the relative SNR of each channel follows the Gamma distribution $\gamma_j \backsim \mathcal{G}(m,1/m),~j=1,\ldots, N$ for a unit average fading power, and same scale and shaping parameters for all the channels. As the signals are coherently combined at the BS following MRC scheme, the aggregate SNR ($\gamma$) at the BS is the sum of the SNRs received at each BS antenna. Given that the sum of $N$ Gamma random variables is also a Gamma distributed random variable, the aggregate SNR $\gamma$ $\backsim \mathcal{G}(Nm,1/m)$}. \textcolor{black}{The effective SINR is expressed as 
\begin{equation}
    \gamma_{\text{eff}} = \frac{\Omega P \| \textbf{h}\|^2/N}{P_r\| \textbf{g}\|^2 + \sigma^2} = \frac{\hat{\gamma}}{\gamma_z + 1},
\end{equation}}
\textcolor{black}{where $\|\textbf{h}\|^2 = \sum_{n=1}^{N}|h_n|^2$, $P_r$ is the power of a single interferer, $\|\textbf{g}\|^2 = \sum_{n=1}^{M_z}|g_n|^2$, $\gamma$ is the updated SNR of the desired signal, and $\gamma_z$ is the SNR of the aggregate interference. According to the literature and with the assumption of rich scattering interference, a good choice for each interferer model is Rayleigh fading. As the average number of interfers is $M_z$, the aggregate SNR distribution follows the Chi-Square distribution with pdf obtained as follows
\begin{equation}\label{eq10}
f_{\gamma_{z}}(x) = \left(\frac{M_z}{\overline{\gamma}_{z}}\right)^{M_z} \frac{x^{M_z-1}}{\Gamma(M_z)}e^{-\frac{M_zx}{\overline{\gamma}_{z}}},
\end{equation}}
\textcolor{black}{where $\overline{\gamma}_z$ is the average SNR of a single interferer. The joint pdf of the outdated and updated SNRs is given by \cite[Eq.~(8)]{int}
\begin{equation}
\begin{split}
f_{\hat{\gamma},\gamma}(x,y) = &\left( \frac{Nm}{\overline{\gamma}}\right)^{Nm+1}\frac{\left(\frac{xy}{\rho}\right)^{\frac{Nm-1}{2}}}{(1-\rho)\Gamma(Nm)}e^{-\left( \frac{x+y}{1-\rho}\right)\frac{Nm}{\overline{\gamma}}}I_{Nm-1}\left(\frac{2Nm\sqrt{\rho xy}}{(1-\rho)\overline{\gamma}}\right),
\end{split}
\end{equation}}
where $I_{\nu}(\cdot)$ is the $\nu$-th order modified Bessel function of the first kind and $\overline{\gamma}$ is the average SNR. \textcolor{black}{Assuming that the UE selects the BS of rank $k$ conditioned on the correlation factor $\rho$ and refering to \cite[Eq.~(28)]{int}, the pdf of the effective SINR $\gamma_{\text{eff}}$ can be expressed as
\begin{equation}\label{pdf}
\begin{split}
f_{\gamma_{\text{eff}}}(x) =& \sum_{n=0}^{k-1}\sum_{i=0}^{j(Nm-1)}\sum_{v=0}^{i}\sum_{u=0}^{Nm+v}
{M \choose k}{k-1 \choose n}{i \choose v}{Nm+v \choose u}\frac{\Gamma(Nm+i)\Gamma(u+M_z)}{\Gamma(M_z)\Gamma(Nm)\Gamma(v+Nm)}\\&\times~\Phi(i,j,Nm-1)\frac{(-1)^nk\beta_z^{M_z}\rho^v(1-\rho)^{i-v}[1+j(1-\rho)]^{M_z+u-v-Nm-1}\overline{\gamma}^{M_z+u-1}}{[Nm(1+j)+(1+j(1-\rho))\overline{\gamma}\beta_z]^{M_z+u-1}}\\&\times~\left(\frac{Nm}{\overline{\gamma}} \right)^{Nm+v}x^{Nm+v-1}e^{-\frac{Nm(1+j)x}{(1+j(1-\rho))\overline{\gamma}}},
\end{split}
\end{equation}}
\textcolor{black}{where the coefficients $\Phi(i,j,m)$ are defined recursively as $(\sum_{i=0}^m\frac{x^i}{i!})^j = \sum_{i=0}^{j(m-1)}\Phi(i,j,m)x^i$, $\Phi(i,j,m)$ = $\sum_{n=n_1}^{n_2}\frac{\Phi(n_1,j-1,m)}{(i-n_1)!}x^i$, $n_1 = \max(0,i-m)$, $n_2 = \min(i,(j-1)(m-1))$, and $\beta_z = \frac{M_z}{\overline{\gamma}_z}$. After using \cite[Eq.~(3.351.1)]{64} to integrate (\ref{pdf}), the cdf of the effective SINR is given by
\begin{equation}\label{cdf}
\begin{split}
&F_{\gamma_{\text{eff}}}(x) =  \sum_{n=0}^{k-1}\sum_{i=0}^{j(Nm-1)}\sum_{v=0}^{i}\sum_{u=0}^{Nm+v}
{M \choose k}{k-1 \choose n}{i \choose v}{Nm+v \choose u}\frac{\Gamma(Nm+i)\Gamma(u+M_z)}{\Gamma(M_z)\Gamma(Nm)}\\&\times~\Phi(i,j,Nm-1)\frac{(-1)^nk\beta_z^{M_z}\rho^v(1-\rho)^{i-v}[1+j(1-\rho)]^{M_z+u-2}\overline{\gamma}^{M_z+Nm+u+v-q-1}}{[Nm(1+j)+(1+j(1-\rho))\overline{\gamma}\beta_z]^{M_z+u-1}}\left(\frac{Nm}{\overline{\gamma}} \right)^{Nm+v}\\&\times~\frac{1}{(Nm(1+j))^{Nm+v-1}}  \left[ 1 - e^{-\frac{Nm(1+j)x}{(1+j(1-\rho))\overline{\gamma}}}\sum_{q=0}^{Nm+v-1}\left(\frac{Nm(1+j)}{[1+j(1-\rho)]\overline{\gamma}}  \right)^q\frac{x^q}{q!} \right],
\end{split}    
\end{equation}}
\textcolor{black}{Using \cite[Eq.~(3.351.3)]{64}, the $p$-th moment of the effective SINR is expressed as follows
\begin{equation}
\begin{split}
\mathbb{E}_{\gamma_{\text{eff}}}[\gamma^p] =& \sum_{n=0}^{k-1}\sum_{i=0}^{j(Nm-1)}\sum_{v=0}^{i}\sum_{u=0}^{Nm+v}
{M \choose k}{k-1 \choose n}{i \choose v}{Nm+v \choose u}\frac{\Gamma(Nm+i)\Gamma(u+M_z)}{\Gamma(M_z)\Gamma(Nm)\Gamma(v+Nm)}\\&\times~\Phi(i,j,Nm-1)\frac{(-1)^nk\beta_z^{M_z}\rho^v(1-\rho)^{i-v}[1+j(1-\rho)]^{M_z+u+p+1}\overline{\gamma}^{M_z+Nm+u+v+p+1}}{[Nm(1+j)+(1+j(1-\rho))\overline{\gamma}\beta_z]^{M_z+u-1}}\\&\times~\left(\frac{Nm}{\overline{\gamma}} \right)^{Nm+v}\frac{\Gamma(Nm+v+p)}{(Nm(1+j))^{Nm+p+v+2}},
\end{split}    
\end{equation}}
\begin{rem}
\textcolor{black}{Note that the moment of the effective SINR is useful to derive the low SNR expansion, and the approximation of the achievable rate which will be detailed later.}
\end{rem}
\subsection{Achievable Rate}
The ergodic achievable rate $\mathcal{C}_1$ of the cellular network, expressed in nats/s/Hz, is defined as the maximum error-free data rate transferred by the system channel. It can be expressed as follows
\begin{equation}\label{ratec1}
\mathcal{C}_1 \delequal \mathbb{E}_{\gamma_{\text{eff}}}[\log(1+\gamma)],
\end{equation}

After some mathematical manipulations, the achievable rate is derived as follows
\begin{equation}\label{rate}
\begin{split}
\mathcal{C}_1 =& \sum_{n=0}^{k-1}\sum_{i=0}^{j(Nm-1)}\sum_{v=0}^{i}\sum_{u=0}^{Nm+v}
{M \choose k}{k-1 \choose n}{i \choose v}{Nm+v \choose u}\frac{\Gamma(Nm+i)\Gamma(u+M_z)}{\Gamma(M_z)\Gamma(Nm)\Gamma(v+Nm)}\\&\times~\Phi(i,j,Nm-1)\frac{(-1)^nk\beta_z^{M_z}\rho^v(1-\rho)^{i-v}[1+j(1-\rho)]^{M_z+u-1}\overline{\gamma}^{M_z+Nm+u+v-1}}{[Nm(1+j)+(1+j(1-\rho))\overline{\gamma}\beta_z]^{M_z+u-1}}\\&\times~\left(\frac{(Nm)^2(j+1)}{\overline{\gamma}} \right)^{Nm+v}G_{3,2}^{1,3} \Bigg(\frac{[1+j(1-\rho)]\overline{\gamma}}{Nm(j+1)} ~\bigg|~\begin{matrix} 1-Nm-v,1,1 \\ 1,0 \end{matrix} \Bigg),
\end{split}
\end{equation}

where $G_{p,q}^{m,n}(\cdot)$ is the Meijer-$G$ function. 
\begin{proof}
The proof of (\ref{rate}) is provided in Appendix A.
\end{proof}
\textcolor{black}{At low SNR, the achievable rate can be expanded as follows
\begin{equation}
\mathcal{C}_1 \cong \mathbb{E}_{\gamma_{\text{eff}}}[\gamma].    
\end{equation}}
\textcolor{black}{In addition, we can derive the Jensen's upper bound for the achievable rate as 
\begin{equation}
\mathcal{C}_1 \leq \log(1+\mathbb{E}_{\gamma_{\text{eff}}}[\gamma]).    
\end{equation}}
\textcolor{black}{The first moment is very useful as we argued earlier since it yields the derivation of the low SNR approximation and the upper bound of the achievable rate.}
\subsection{Blockage Model}
\textcolor{black}{MmWaves communications are very sensitive to blockage, where relative models have been extensively studied in the literature \cite{e5,e6,e7}. The proposed models basically depend on the geometry of the objects, and the density $\mu$ in a given area (urban, suburban, and rural). In this work, we will focus on the following blockage models.
\begin{equation}{\label{eq34}}
p_{\text{los}} = e^{-\frac{d}{\mu}},    
\end{equation}}
\textcolor{black}{where $p_{\text{los}}$ is the probability of LOS, and $d$ is the distance between the TX and RX. This model is called the exponential blockage model. According to 3GPP, $\mu = 200$ m in a suburban area, and $\mu = 63$ m in an urban area \cite{rap}}.

\section{FSO Backhauling Analysis}
\subsection{Channel Model}
The FSO part consists of three components $I_a, I_l$, and $I_p$ which are the turbulence-induced fading, the path loss, and the pointing errors fading, respectively. The channel gain $I_z$ of the FSO between the BS and the data center can be expressed as follows
\begin{equation}\label{eq:9}
I_z = I_a I_l I_p. 
\end{equation}

Using the Beers-Lambert law, the path loss can be expressed as follows
\textcolor{black}{\begin{equation}\label{pathloss}
    I_l = \frac{\pi a^2}{(\theta L_2)^2}\exp(-\sigma L_2),
\end{equation}}
\textcolor{black}{where $a$ is the radius of the receiver aperture, $\theta$ is the receive beam divergence, $L_2$ is the optical link distance between the BS and the data center, and $\sigma$ is the weather attenuation coefficient}.
The pointing errors $I_p$ made by Jitter can be given as \cite[Eq.~(9)]{ln1}
\begin{equation}\label{pointing}
I_p = A_0 \exp\left(-\frac{2R^2}{\omega^2_{z_{eq}}} \right),
\end{equation}

where $\omega_{z_{eq}}$ is the equivalent beam waist. Assuming that the radial displacement $R$ of the beam at the detector follows the Rayleigh distribution, the pdf of the pointing errors can be expressed as follows
\begin{equation}\label{eq:12}
f_{I_p}(I_p) = \frac{\xi^2}{A_0^{\xi^2}}I^{\xi^2-1}_p~,~~0\leq I_p\leq A_0
\end{equation}

The pointing errors coefficient can be expressed in terms of the Jitter standard deviation ($\sigma_s$) and the equivalent beam waist as follows
\begin{equation}\label{eq:13}
\xi = \frac{\omega_{z_{eq}}}{2\sigma_{\text{s}}}.
\end{equation}

We can also relate $\omega_{z_{eq}}$ with the beam width $\omega_z$ of the Gaussian laser beam at the distance $L_2$ as follows
\begin{equation}\label{eq:14}
\omega^2_{z_{eq}} = \frac{\omega^2_{L_2}\sqrt{\pi}\text{erf}(v)}{2v\exp(-v^2)},
\end{equation}

where $v =\frac{\sqrt{\pi}a}{\sqrt{2}\omega_{L_2}}$, and \text{erf}($\cdot$) is the error function. The fraction of the collected power at the relay is $A_0 = |\text{erf}(v)|^2$. The Gaussian beam waist can be defined as
\begin{equation}\label{eq:16}
\begin{split}
\omega_{z} = \omega_0\sqrt{(\Theta_0 + \Lambda_0)(1 + 1.63~\sigma_{\text{Rytov}}^{12/5}\Lambda_1)},
\end{split}
\end{equation}

where $\Theta_0 = 1 - \frac{L_2}{F_0},~\Lambda_0 = \frac{\lambda_2L_2}{\pi \omega_0^2},~\Lambda_1 = \frac{\Lambda_0}{\Theta_0^2 + \Lambda^2_0}$, and $\sigma_{\text{Rytov}}^2$ is the Rytov variance given by \cite[Eq.~(15)]{49}
\begin{equation}\label{eq:17}
\sigma^2_{\text{Rytov}} = 1.23~C^2_nL_2^{11/6}\left(\frac{2\pi}{\lambda_2} \right)^{7/6},
\end{equation}

where $\lambda_2$ is the wavelength of FSO laser beam, $F_0$ is the radius of the curvature, and $C_n^2$ is the refractive index of the medium. The turbulence-induced fading $I_a$ is modeled by the Double Generalized Gamma and can be expressed as the product of two independent random variables $I_x$ and $I_y$ describing the large-scale and small-scale fluctuations, respectively. $I_x$ and $I_y$ each follows the generalized gamma distribution $I_x  \backsim GG(\alpha_1, m_1, \Omega_1)$ and $I_y  \backsim GG(\alpha_2, m_2, \Omega_2)$, where $m_1$ and $m_2$ are the shaping parametes defining the atmospheric turbulence fading. Moreover, $\alpha_1, \alpha_2, \Omega_1$, and $\Omega_2$ are defined using the variance of the small and large scale fluctutaions from \cite[Eqs.~(8.a), (8.b), and (9)]{dgg}. Thereby, the pdf of the turbulence-induced fading $I_a$ can be given by \cite[Eq.~(4)]{dgg}
\begin{align}\begin{split} 
{f_{I_a}}\left({I_a} \right) = \frac{\alpha _{2}{p^{{m_2} + \frac{1}{2}}}{q^{{m_1} - \frac{1}{2}}}{\left({2\pi } \right)}^{1 - \frac{p + q}{2}}}{\Gamma \left({m_1} \right)\Gamma \left({m_2} \right){I_a}}G_{p + q,0}^{0,p + q}\left(\frac{{p^p}{q^q}\Omega _1^q\Omega _2^p}{{m_1^q}{m_2^p}I_a^{\alpha_2p}} \left\vert\begin{array}{c} {\Delta \left({q{:}1 - {m_1}} \right)},{\Delta \left({p{:}1 - {m_2}} \right)}\\ {-}\end{array}\right. \!\!\right),\end{split}\end{align}
where $p$ and $q$ are positive integers satisfying $\frac{p}{q} = \frac{\alpha_1}{\alpha_2}$ and $\Delta(j~;~x) \delequal \frac{x}{j}, \ldots, \frac{x + j - 1}{j}$. In case of the heterodyne detection, the average SNR $\mu_1$ is given by $\mu_1 = \frac{\eta\e{I_z}}{\sigma^2_2}$. Regarding the IM/DD detection, the average electrical SNR $\mu_2$ is given by $\mu_2 = \frac{(\eta\e{I_z})^2}{\sigma^2_2}$ while the instantaneous optical SNR is $\gamma_{r} = \frac{(\eta I^2_z)}{\sigma^2_2}$. Unifying the two detection schemes and applying the transformation of the random variable $\gamma_{r} = \frac{(\eta I_z)^r}{\sigma_2^2}$, the unified pdf of the instantaeous SNR $\gamma_{r}$ can be expressed as follows
\begin{equation}\label{eq:19}
\begin{split}
f_{\gamma_{r}}(\gamma) = \frac{\xi^2p^{m_2-\frac{1}{2}}q^{m_1-\frac{1}{2}}(2\pi)^{1-\frac{p+q}{2}}}{r\Gamma(m_1)\Gamma(m_2)\gamma} G_{p+q+\alpha_2p,\alpha_2p}^{0,p+q+\alpha_2p} \Bigg(\left(\frac{p\Omega_1}{m_2}\right)^p \left(\frac{q\Omega_2}{m_1}\right)^q \left(\frac{\mu_r(A_0 I_l)^{r}}{\gamma}\right)^{\frac{\alpha_2p}{r}} ~\bigg|~\begin{matrix} \kappa_1 \\ \kappa_2 \end{matrix} \Bigg),
\end{split}
\end{equation}
where $\sigma_2^2$, $\eta$ are the noise at the Rx data center and the electrical-to-optical conversion coefficient, respectively. The parameter $r$ takes two values 1 and 2 standing for heterodyne and IM/DD, respectively. The vectors $\kappa_1 =[\Delta(\alpha_2p:1-\xi^2),~\Delta(q:1 - m_1),~\Delta(p:1 - m_2)] $, and $\kappa_2 = [\Delta(p:1 - m_2),~\kappa_2 = \Delta(\alpha_2p:-\xi^2)]$. The average SNR $\overline{\gamma}_r$\footnote[1]{The average SNR $\overline{\gamma}_r$ is defined as $\overline{\gamma}_r = \eta^r\e{I_{z}^r}/\sigma_{2}^2$, while the average electrical SNR $\mu_r$ is given by $\mu_r = \eta^r\e{I_{z}}^r/\sigma_{2}^2$. Therefore, the relation between the average SNR and the average electrical SNR is trivial given that $ \frac{\e{I^2_{z}}}{\e{I_{z}}^2} = \sigma^2_{\text{si}} + 1$, where $\sigma^2_{\text{si}}$ is the scintillation index \cite{scin}.} can be expressed as
\begin{align}
 \overline{\gamma}_r = \frac{\e{I^r_{z}}}{\e{I_{z}}^r}\mu_r,   
\end{align}

where $\mu_r$ is the average electrical SNR given by
\begin{equation}
    \mu_r = \frac{\eta^r\e{I_{z}}^r}{\sigma_{2}^2}.
\end{equation}
After integrating Eq.~(19), the cdf of the instantaneous SNR $\gamma_{2(m)}$ can be expressed as follows
\begin{equation}\label{eq:21}
\begin{split}
F_{\gamma_{r}}(\gamma) = \frac{\xi^2p^{m_2-\frac{3}{2}}q^{m_1-\frac{1}{2}}(2\pi)^{1-\frac{p+q}{2}}}{\alpha_2\Gamma(m_1)\Gamma(m_2)}  G_{p+q+2\alpha_2p,2\alpha_2p}^{\alpha_2p,p+q+\alpha_2p} \Bigg(\left(\frac{p\Omega_1}{m_2}\right)^p \left(\frac{q\Omega_2}{m_1}\right)^q \left(\frac{\mu_r(A_0 I_l)^{r}}{\gamma}\right)^{\frac{\alpha_2p}{r}}\bigg|\begin{matrix} \kappa_3 \\ \kappa_4 \end{matrix} \Bigg),
\end{split}
\end{equation}

where $\kappa_3 = [\Delta(\alpha_2p:1-\xi^2),~\Delta(q:1 - m_1),~\Delta(p:1 - m_2),~[1]_{\alpha_2p}]$, $\kappa_4 = [[0]_{\alpha_2p},~\Delta(\alpha_2p:-\xi^2)]$, and $[x]_j$ is defined as the vector of length $j$ and its components are equal to $x$.

After changing the variable of the integration ($x = \gamma^{-\frac{\alpha_2p}{r}
}$) and applying the following identity \cite[Eq.~(2.24.2.1)]{62}, the $t$-th moment of the optical SNR can be derived as follows
\begin{equation}\label{eq:22}
\begin{split}
\e{\gamma_{r}^t} = \ddfrac{\xi^2p^{m_2-1}q^{m_1-\frac{1}{2}}(2\pi)^{1-\frac{p+q}{2}}\chi^{t\left[\frac{r}{\alpha_2p}-1\right]-1}}{\Gamma(m_1)\Gamma(m_2)\prod_{j=1}^{\alpha_2p}\Gamma\left(t\left[\frac{r}{\alpha_2p}-1\right]-\kappa_{2,j}\right)}
\frac{\prod_{j=1}^{p+q+\alpha_2p}\Gamma\left(t\left[\frac{r}{\alpha_2p}-1\right]-\kappa_{1,j}\right)}{\prod_{j=p+q+\alpha_2p+1}^{p+q+2\alpha_2p}\Gamma\left(t\left[\frac{r}{\alpha_2p}-1\right]-\kappa_{1,j}\right)},
\end{split}
\end{equation}
where $\chi = \left(\frac{p\Omega_1}{m_2}\right)^p \left(\frac{q\Omega_2}{m_1}\right)^q (A_0 I_l)^{\alpha_2p}\mu_r^{\frac{\alpha_2p}{r}}$.
\subsection{Non linear HPA Models Analysis}
Since the distortion created by the HPA is not linear and so the analysis will be somewhat complex, we refer to the Bussgang linearization theory to linearize the distortion. This theory states that the output of the non linear HPA circuit is a function of the linear scale parameter $\zeta$ of the input signal and a non linear distortion $\varsigma$ uncorrelated with the input signal and modeled as a complex Gaussian random variable $\varsigma \backsim \mathcal{CN} (0,~\sigma^2_{\varsigma})$. According to \cite{28,ofdm}, the parameters $\zeta$ and $\sigma^2_{\varsigma}$ for SEL are given by \cite[Eq.~(17)]{27}
\begin{equation}
    \zeta = 1 - \exp\left(-\frac{A^2_{\text{sat}}}{\sigma^2_{r}}\right) + \frac{\sqrt{\pi} A_{\text{sat}}}{2\sigma^2_{r}}~\text{erfc}\left(\frac{A_{\text{sat}}}{\sigma_{r}}\right).
\end{equation}
\begin{equation}
\sigma^2_{\varsigma} = \sigma^2_{r}~\left[1 - \exp\left(-\frac{A^2_{\text{sat}}}{\sigma^2_{r}}\right) - \zeta^2 \right].    
\end{equation}

For TWTA, $\zeta$ and $\sigma^2_{\varsigma}$ are given by \cite[Eq.~(18)]{27}
\begin{equation}
\zeta = \frac{A^2_{\text{sat}}}{\sigma^2_{r}}\left[1 + \frac{A^2_{\text{sat}}}{\sigma^2_{r}}\exp\left(\frac{A^2_{\text{sat}}}{\sigma^2_{r}}\right)  + \text{Ei}\left(-\frac{A^2_{\text{sat}}}{\sigma^2_{r}} \right)   \right].
\end{equation}
\begin{equation}
\sigma^2_{\varsigma} = -\frac{A^4_{\text{sat}}}{\sigma^2_{r}}\left[\left(1 + \frac{A^2_{\text{sat}}}{\sigma^2_{r}}\right)\exp\left(\frac{A^2_{\text{sat}}}{\sigma^2_{r}}\right)\text{Ei}\left(-\frac{A^2_{\text{sat}}}{\sigma^2_{r}} \right) + 1 \right]-\sigma^2_{r}\zeta^2.    
\end{equation}

\textcolor{black}{Assuming a unit smoothness factor, the SSPA parameters are derived by \cite[Eq.~(18)]{28}}
\textcolor{black}{\begin{equation}
\zeta = \frac{A_{\text{sat}}}{2\sigma_r}\left[\frac{2A_{\text{sat}}}{\sigma_r}-\sqrt{\pi}\text{erfc}\left( \frac{A_{\text{sat}}}{\sigma_r}\right)\exp\left(-\frac{A^2_{\text{sat}}}{\sigma^2_{r}}\right)\left(\frac{2A^2_{\text{sat}}}{\sigma^2_{r}}-1  \right)\right].   
\end{equation}
\begin{equation}
\sigma^2_{\varsigma} = \sigma_r^2\left[\frac{A^2_{\text{sat}}}{\sigma^2_{r}} \left(1+ \frac{A^2_{\text{sat}}}{\sigma^2_{r}}\exp\left(\frac{A^2_{\text{sat}}}{\sigma^2_{r}}\right)\text{Ei}\left(-\frac{A^2_{\text{sat}}}{\sigma^2_{r}} \right) \right)-\zeta^2 \right],    
\end{equation}}

\textcolor{black}{where $A_{\text{sat}}$, $\sigma^2_{r}$, $\text{erfc}(\cdot)$, and $\text{Ei}(\cdot)$ are the input saturation amplitude of the power amplifier, the mean power of the signal at the output of the gain block, the complementary error function, and the exponential integral function, respectively.}
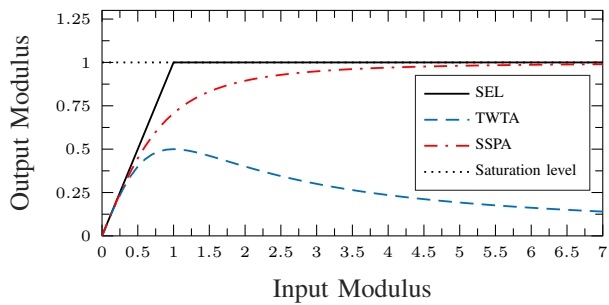
\begin{figure}[H]
\centering
\setlength\fheight{3cm}
\setlength\fwidth{7cm}
%
%
\begin{tikzpicture}

\begin{axis}[%
width=0.951\fwidth,
height=\fheight,
at={(0\fwidth,0\fheight)},
scale only axis,
xmin=0,
xmax=7,
xtick = {0,.5,1,1.5,2,2.5,3,3.5,4,4.5,5,5.5,6,6.5,7},
xlabel style={font=\color{white!15!black}},
xlabel={\small{Input Modulus}},
ymin=0,
ymax=1.3,
ytick = {0,.25,.5,.75,1,1.25,1.5},
ylabel style={font=\color{white!15!black}},
ylabel={\small{Output Modulus}},
axis background/.style={fill=white},
legend style={at={(.8,0.2)}, anchor=south,legend cell align=left, align=left, draw=white!15!black}
]
\addplot [color=black, line width=0.7pt]
  table[row sep=crcr]{%
0	0\\
0.1	0.1\\
0.2	0.2\\
0.3	0.3\\
0.4	0.4\\
0.5	0.5\\
0.6	0.6\\
0.7	0.7\\
0.8	0.8\\
0.9	0.9\\
1	1\\
1.1	1\\
1.2	1\\
1.3	1\\
1.4	1\\
1.5	1\\
1.6	1\\
1.7	1\\
1.8	1\\
1.9	1\\
2	1\\
2.1	1\\
2.2	1\\
2.3	1\\
2.4	1\\
2.5	1\\
2.6	1\\
2.7	1\\
2.8	1\\
2.9	1\\
3	1\\
3.1	1\\
3.2	1\\
3.3	1\\
3.4	1\\
3.5	1\\
3.6	1\\
3.7	1\\
3.8	1\\
3.9	1\\
4	1\\
4.1	1\\
4.2	1\\
4.3	1\\
4.4	1\\
4.5	1\\
4.6	1\\
4.7	1\\
4.8	1\\
4.9	1\\
5	1\\
5.1	1\\
5.2	1\\
5.3	1\\
5.4	1\\
5.5	1\\
5.6	1\\
5.7	1\\
5.8	1\\
5.9	1\\
6	1\\
6.1	1\\
6.2	1\\
6.3	1\\
6.4	1\\
6.5	1\\
6.6	1\\
6.7	1\\
6.8	1\\
6.9	1\\
7	1\\
};
\addlegendentry{\tiny{SEL}}

\addplot [color=NavyBlue, dash pattern={on 5pt off 3pt on 0pt off 0pt}, line width=0.7pt]
  table[row sep=crcr]{%
0	0\\
0.1	0.099009900990099\\
0.2	0.192307692307692\\
0.3	0.275229357798165\\
0.4	0.344827586206897\\
0.5	0.4\\
0.6	0.441176470588235\\
0.7	0.469798657718121\\
0.8	0.48780487804878\\
0.9	0.497237569060773\\
1	0.5\\
1.1	0.497737556561086\\
1.2	0.491803278688525\\
1.3	0.483271375464684\\
1.4	0.472972972972973\\
1.5	0.461538461538462\\
1.6	0.449438202247191\\
1.7	0.437017994858612\\
1.8	0.424528301886792\\
1.9	0.412147505422993\\
2	0.4\\
2.1	0.388170055452865\\
2.2	0.376712328767123\\
2.3	0.365659777424483\\
2.4	0.355029585798817\\
2.5	0.344827586206897\\
2.6	0.335051546391753\\
2.7	0.325693606755127\\
2.8	0.316742081447964\\
2.9	0.308182784272051\\
3	0.3\\
3.1	0.292177191328935\\
3.2	0.284697508896797\\
3.3	0.277544154751892\\
3.4	0.270700636942675\\
3.5	0.264150943396226\\
3.6	0.257879656160458\\
3.7	0.251872021783526\\
3.8	0.246113989637306\\
3.9	0.240592227020358\\
4	0.235294117647059\\
4.1	0.230207748455924\\
4.2	0.225321888412017\\
4.3	0.220625962031811\\
4.4	0.216110019646365\\
4.5	0.211764705882353\\
4.6	0.207581227436823\\
4.7	0.203551320918146\\
4.8	0.199667221297837\\
4.9	0.195921631347461\\
5	0.192307692307692\\
5.1	0.188818955942244\\
5.2	0.185449358059914\\
5.3	0.182193193537298\\
5.4	0.179045092838196\\
5.5	0.176\\
5.6	0.173053152039555\\
5.7	0.170200059719319\\
5.8	0.16743648960739\\
5.9	0.164758447361072\\
6	0.162162162162162\\
6.1	0.1596440722324\\
6.2	0.157200811359026\\
6.3	0.154829196362743\\
6.4	0.152526215443279\\
6.5	0.15028901734104\\
6.6	0.148114901256733\\
6.7	0.146001307474395\\
6.8	0.143945808636749\\
6.9	0.14194610162518\\
7	0.14\\
};
\addlegendentry{\tiny{TWTA}}

\addplot [color=red, dash pattern={on 5pt off 3pt on 1pt off 3pt} , line width=0.7pt]
  table[row sep=crcr]{%
0	0\\
0.1	0.0995037190209989\\
0.2	0.196116135138184\\
0.3	0.287347885566345\\
0.4	0.371390676354104\\
0.5	0.447213595499958\\
0.6	0.514495755427527\\
0.7	0.573462344363328\\
0.8	0.624695047554424\\
0.9	0.66896473162245\\
1	0.707106781186547\\
1.1	0.739940073395944\\
1.2	0.768221279597376\\
1.3	0.7926239891046\\
1.4	0.813733471206735\\
1.5	0.832050294337844\\
1.6	0.847998304005088\\
1.7	0.86193421515777\\
1.8	0.874157276121538\\
1.9	0.884918222381982\\
2	0.894427190999916\\
2.1	0.90286051882393\\
2.2	0.910366477462605\\
2.3	0.917070056253235\\
2.4	0.923076923076923\\
2.5	0.928476690885259\\
2.6	0.93334560620306\\
2.7	0.937748760723704\\
2.8	0.941741911594837\\
2.9	0.945372981626272\\
3	0.948683298050514\\
3.1	0.951708617760551\\
3.2	0.95447997803503\\
3.3	0.957024404433473\\
3.4	0.959365501571271\\
3.5	0.961523947640823\\
3.6	0.963517909629941\\
3.7	0.965363393028266\\
3.8	0.967074537262646\\
3.9	0.968663866044045\\
4	0.970142500145332\\
4.1	0.971520338783129\\
4.2	0.972806214685367\\
4.3	0.974008027039196\\
4.4	0.97513285579146\\
4.5	0.976187060183953\\
4.6	0.977176363922801\\
4.7	0.978105928984835\\
4.8	0.978980419737605\\
4.9	0.979804058780407\\
5	0.98058067569092\\
5.1	0.981313749677157\\
5.2	0.982006446980647\\
5.3	0.982661653748471\\
5.4	0.98328200498446\\
5.5	0.983869910099907\\
5.6	0.984427575508482\\
5.7	0.984957024646314\\
5.8	0.985460115744348\\
5.9	0.985938557634464\\
6	0.986393923832144\\
6.1	0.986827665105534\\
6.2	0.987241120712647\\
6.3	0.987635528464463\\
6.4	0.988012033751101\\
6.5	0.988371697650617\\
6.6	0.988715504224767\\
6.7	0.989044367093028\\
6.8	0.989359135364853\\
6.9	0.989660599000355\\
7	0.989949493661167\\
};
\addlegendentry{\tiny{SSPA}}

\addplot [color=black, dotted, line width=0.7pt]
  table[row sep=crcr]{%
0	1\\
0.1	1\\
0.2	1\\
0.3	1\\
0.4	1\\
0.5	1\\
0.6	1\\
0.7	1\\
0.8	1\\
0.9	1\\
1	1\\
1.1	1\\
1.2	1\\
1.3	1\\
1.4	1\\
1.5	1\\
1.6	1\\
1.7	1\\
1.8	1\\
1.9	1\\
2	1\\
2.1	1\\
2.2	1\\
2.3	1\\
2.4	1\\
2.5	1\\
2.6	1\\
2.7	1\\
2.8	1\\
2.9	1\\
3	1\\
3.1	1\\
3.2	1\\
3.3	1\\
3.4	1\\
3.5	1\\
3.6	1\\
3.7	1\\
3.8	1\\
3.9	1\\
4	1\\
4.1	1\\
4.2	1\\
4.3	1\\
4.4	1\\
4.5	1\\
4.6	1\\
4.7	1\\
4.8	1\\
4.9	1\\
5	1\\
5.1	1\\
5.2	1\\
5.3	1\\
5.4	1\\
5.5	1\\
5.6	1\\
5.7	1\\
5.8	1\\
5.9	1\\
6	1\\
6.1	1\\
6.2	1\\
6.3	1\\
6.4	1\\
6.5	1\\
6.6	1\\
6.7	1\\
6.8	1\\
6.9	1\\
7	1\\
};
\addlegendentry{\tiny{Saturation level}}

\end{axis}
\end{tikzpicture}%
    \caption{\textcolor{black}{AM/AM characteristics of SEL, TWTA, and SSPA with unit smoothness factor.}}
\end{figure}

\textcolor{black}{Further details about the derivation of the AM/AM of SEL, TWTA, and SSPA are provided by \cite{28}. 
\subsection{Effective optical signal-to-noise-plus-distortion ratio (SNDR)}
As the BS amplifies the re-encoded signal with an amplification gain $G$, the non linear HPA distortion factor, $\kappa$, is given by}
\begin{equation}
\kappa = 1 + \frac{\sigma^2_{\varsigma}}{\zeta^2G^2\sigma_1^2}.
\end{equation}

Using \cite[Eq.~(12)]{aggregate}, the effective (non ideal hardware) optical SNDR is expressed as
\begin{equation}
\gamma_{\text{ni}} = \frac{\gamma_r}{(\kappa-1)\gamma_r+1}.    
\end{equation}

\textcolor{black}{Consequently, the cdf of the SNDR is derived as follows
\begin{equation}\label{cdf-sndr}
F_{\gamma_{\text{ni}}}(x) = \left\{
        \begin{array}{ll}
            F_{\gamma_r}\left(\frac{x}{1-(\kappa-1)x}\right) & \quad \text{if}~ x < \frac{1}{\kappa-1}, \\
            1 & \quad \text{otherwise},
        \end{array}
    \right.
\end{equation}}
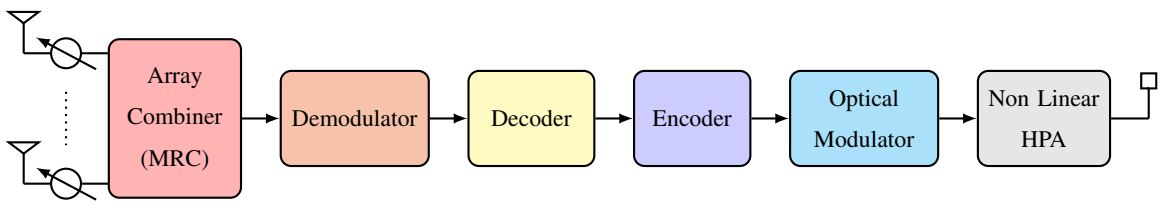
\begin{figure}[H]
\centering
\setlength\fheight{5cm}
\setlength\fwidth{20cm}

\usetikzlibrary{shapes.misc,shapes.geometric,shapes.symbols,positioning,shadings,automata}
\tikzstyle{XORgate} = [draw,circle]
\newcommand{\antena}{--++(3mm,0)--++(30:5mm)--++(-90:5mm)--++(150:5mm);}
\newcommand{\suma}{\Large$+$}

\definecolor{mycolor1}{rgb}{1.00000,0.00000,1.00000}%
\definecolor{orange}{rgb}{0.9100,0.4100,0.1700}

\pgfdeclarelayer{background}
\pgfdeclarelayer{foreground}
\pgfsetlayers{background,main,foreground}
\usetikzlibrary{shapes,arrows}
\newcommand{\mx}[1]{\mathbf{\bm{#1}}} 
\newcommand{\vc}[1]{\mathbf{\bm{#1}}} 

\tikzstyle{baseband} = [draw, text width=3.3em, fill=orange!50, text centered,
    minimum height=12em, rounded corners,font = \footnotesize]

\tikzstyle{chain}=[draw, fill=red!30, text width=3.5em, 
    text centered, minimum height=5em, rounded corners, font = \footnotesize]
\tikzstyle{decoder}=[draw, fill=orange!40, text width=4em, 
    text centered, minimum height=3em, rounded corners, font = \footnotesize] 
    
\tikzstyle{decoder1}=[draw, fill=yellow!30, text width=3.3em, 
    text centered, minimum height=3em, rounded corners, font = \footnotesize] 
\tikzstyle{decoder2}=[draw, fill=cyan!30, text width=4em, 
    text centered, minimum height=3em, rounded corners, font = \footnotesize]  
    
\tikzstyle{decoder3}=[draw, fill=gray!20, text width=3.5em, 
    text centered, minimum height=3em, rounded corners, font = \footnotesize]      
    
\tikzstyle{encoder}=[draw, fill=blue!20, text width=3em, 
    text centered, minimum height=3em, rounded corners, font = \footnotesize]

\tikzstyle{phase} = [draw,circle,radius=1cm,fill=none]
\tikzstyle{position} = [draw=none,fill=none]
    
\tikzstyle{ann} = [above, text width=5em,font = \footnotesize]

\tikzstyle{center1}=[draw=white,fill=white!20,font = \footnotesize]

\tikzstyle{node1} = [sensor, text width=1.3em, fill=cyan!50, 
    minimum height=10em, rounded corners,font = \footnotesize]
    
\tikzstyle{node2} = [sensor, text width=1.3em, fill=yellow!30, 
    minimum height=10em, rounded corners,font = \footnotesize]  
    
\tikzstyle{node3} = [sensor, text width=2.8em, fill=blue!40, 
    minimum height=7em, rounded corners,font = \footnotesize]           
   
\tikzstyle{circ} = [draw,circle,radius=0.5cm,fill=red]
\tikzstyle{arr} = [draw,circle,radius=0.5cm,fill=red]

\def\blockdist{2.3}
\def\edgedist{2.5}
\def\antenna{%
    -- +(0mm,4.0mm) -- +(2mm,5.5mm) -- +(-2mm,5.5mm) -- +(0mm,4.0mm)
}

\def\aperture{%
    -- +(0mm,4.0mm) -- +(1mm,4mm) -- +(1mm,6mm) -- +(-1mm,6mm) -- +(-1mm,4mm)-- +(0mm,4mm)
} 

\usetkzobj{all}
\begin{tikzpicture}[thick]
\node (pos0) [position] {~};


\node [right = 1mm of pos0](chain3) [chain]{Array Combiner (MRC)};
\node [right = 5mm of chain3](eq) [decoder]{Demodulator};
\node [right = 5mm of eq](dec1) [decoder1]{Decoder};
\node [right = 5mm of dec1](enc) [encoder]{Encoder};
\node [right = 5mm of enc](dec2) [decoder2]{Optical Modulator};
\node [right = 5mm of dec2](dec3) [decoder3]{Non Linear HPA};
\draw[thick] ([xshift=5mm]dec3.east) \aperture;

\draw [solid,thick,-latex] (chain3.east) --  node [right] {} (eq.west); \draw [solid,thick,-latex] (eq.east) --  node [right] {} (dec1.west);
\draw [solid,thick,-latex] (dec1.east) --  node [right] {} (enc.west); 
\draw [solid,thick,-latex] (enc.east) --  node [right] {} (dec2.west);
\draw [solid,thick,-latex] (dec2.east) --  node [right] {} (dec3.west); 
\draw [solid,thick,-] (dec3.east) --  node [right] {} ([xshift=5mm]dec3.east);

\node (pos4) [left = .4cm of chain3,position] {~}; 
\node (phase5) [above = .5cm of pos4,phase] {~}; 
\draw [-latex,thick] ([xshift=4mm]phase5.south) --  node [right] {} ([xshift=-4mm]phase5.north) ; 

\node (phase6) [below = .5cm of pos4,phase] {~}; 
\draw [-latex,thick] ([xshift=4mm]phase6.south) --  node [right] {} ([xshift=-4mm]phase6.north) ; 

\draw [dotted,thick] ([yshift=2mm]pos4.north) --  node [right] {} ([yshift=-2mm]pos4.south);

\draw [solid,thick] (phase5.west) --  node [right] {} ([xshift=-3.5mm]phase5.west);

\draw [solid,thick] (phase6.west) --  node [right] {} ([xshift=-3.5mm]phase6.west);

\draw [solid,thick] (phase5.east) --  node [right] {} ([xshift=3.5mm]phase5.east);
\draw [solid,thick] (phase6.east) --  node [right] {} ([xshift=3.5mm]phase6.east);

\draw[thick] ([xshift=-3.5mm]phase5.west) \antenna;  
\draw[thick] ([xshift=-3.5mm]phase6.west) \antenna;

\end{tikzpicture}
    \caption{\textcolor{black}{Block diagram of the signal processing phases achieved by the BS or the relay RF-to-FSO converter. The mmWaves signal is combined at the Rx array and forwarded by the BS aperture}.}
     \label{node}
\end{figure}
\subsection{Achievable Rate}
The average achievable rate of the FSO backhauling system is expressed as
\begin{equation}\label{rate-fso}
\mathcal{C}_2 = \mathbb{E}_{\gamma_{\text{ni}}}[\log(1+\varpi\gamma)],  
\end{equation}

where $\varpi$ can take the values 1 or $e/2\pi$ for heterodyne or IM/DD, respectively. Due to the presence of the hardware impairments factor, a closed-form expression of the achievable rate is not tractable. Consequently a numerical integration is required to evaluate the exact ergodic rate. Fortunately, we can still derive an approximated expression for the capacity using \cite[Eq.~(35)]{32}
\begin{equation}\label{approx}
\mathbb{E}\left[\log\left(1+\frac{\varphi}{\psi}\right)\right] \cong \log\left(1 + \frac{\mathbb{E}[\varphi]}{\mathbb{E}[\psi]}\right).    
\end{equation}

Although there is no theoretical foundation for (\ref{approx}), yet it still yields an acceptable approximation to the exact expression. We can also characterize the capacity by considering the Jensen's upper bound using the following Theorem.
\begin{thm}
Applying the Jensen's inequality, the upper bound of the achievable rate is expressed as follows
\end{thm}
\begin{equation}\label{jensen}
\mathcal{C}_2 \leq \log\left(1 + \varpi\mathbb{E}_{\gamma_{\text{ni}}}[\gamma]\right).    
\end{equation}

The expectation of $\gamma_{\text{ni}}$ will be numerically evaluated. \textcolor{black}{At high SNR, the achievable rate can be approximated as
\begin{equation}\label{ceiling}
\mathcal{C}_2 \cong \log\left(1 + \frac{\varpi}{\kappa-1}\right).   
\end{equation}}
\textcolor{black}{Expression (\ref{ceiling}) provides a valuable insight. In fact, the capacity converges to a finite ceiling caused by the hardware impairments when the average SNR becomes large. The capacity ceiling cannot be reduced by acting on any system parameters as it is hadware-dependent. Although the ceiling also depends on the detection technique, such impact is still negligible. For ideal hardware ($\kappa = 1$), the capacity ceiling disappears and the rate is not upperbounded for high average SNR}.
\section{Performance Analysis}
\begin{figure}[H]
\centering
\setlength\fheight{6cm}
\setlength\fwidth{12cm}
\input{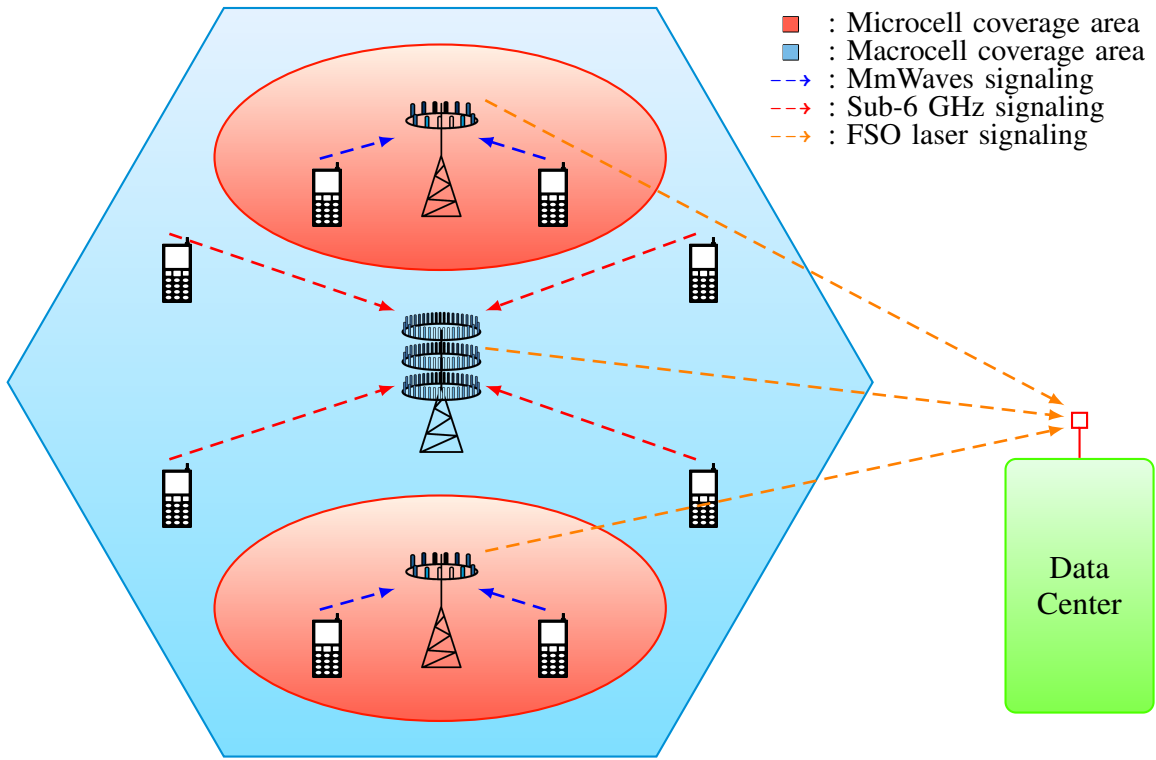}
    \caption{\textcolor{black}{Outdoor heterogenous mmWaves cellular network with FSO backhauling. Sub-6 GHz communications take place within the macrocells (large area) where reliable links mainly require high power to maintain the coverage stability. Inversely, mmWaves signaling is reliable in microcells where the cell area is small, the users density is important and hence high data rate is primarily required. }}
    \label{cellular}
\end{figure}

\textcolor{black}{The system consists of an outdoor heterogenous network where the UE can transmit to either micro or macro BS depending on its serving cell. The transmitted signal undergoes processing by the BS and is then forwarded to the data center. Fig.~\ref{cellular} illustrates the proposed cellular network model with the FSO backhauling. Most importantly, the end-to-end SINDR achieved by the hybrid system is expressed as
\begin{equation}
\gamma_{e2e} = \min\left(\gamma_{\text{eff}},~\gamma_{\text{ni}} \right).
\end{equation}}

\textcolor{black}{Note that such form of the overall SNDR is used in the literature to derive tractable results for amplify-and-forward variable relaying scheme. Yet such approach does not yield exact results for the variable relaying mode, however, it offers exact formulation for the DF scheme which outperforms the amplify-and forward variable/fixed relaying scheme.}
\subsection{Outage Probability}
The outage probability is defined as the probability that the end-to-end SINDR falls below a target threshold $\beta$. It can be generally written as
\begin{equation}\label{out}
p_{\text{out}}(\beta) = \mathbb{P}[\gamma_{\text{e2e}} \leq \beta] = 1 - p_c(\beta).
\end{equation} 

\textcolor{black}{Note that $p_{\text{out}}$ is the cdf of the overall SINDR, and $p_c$ is the coverage probability. Since LOS and NLOS are both considered, the cdf of the SINR relative to the cellular networks is evaluated on average as follows
\begin{equation}
F_{\gamma_{\text{eff}}}(\beta) = p_{\text{los}}F_{\gamma_{\text{eff}}}^{\text{los}}(\beta) + (1-p_{\text{los}}) F_{\gamma_{\text{eff}}}^{\text{nlos}}(\beta).    
\end{equation}}

\textcolor{black}{where $F_{\gamma_{\text{eff}}}^{\text{los}}(\beta)$, and $F_{\gamma_{\text{eff}}}^{\text{nlos}}(\beta)$ are the cdfs of the effective SINR evaluated when the link is LOS, and NLOS, respectively. Consequently, the probability of outage can be expressed in terms of the cdfs of the cellular networks and the FSO bachkauling as follows}

\begin{equation}
p_{\text{out}}(\beta) =  F_{\gamma_{\text{eff}}}(\beta)  +F_{\gamma_{\text{ni}}}(\beta) - F_{\gamma_{\text{eff}}}(\beta)F_{\gamma_{\text{ni}}}(\beta).
\end{equation}

\textcolor{black}{Note that the cellular system achieves a full diverity gain $\mathcal{G} = Nm$ for perfect correlation ($\rho = 1$), otherwise $\mathcal{G} = 1$ for ($\rho <1$). Additionally, we derive an asymptotic high SNR using the Meijer-G expansion of the cdf of $\gamma_{\text{ni}}$ as follows
\begin{equation}\label{eq:40}
\begin{split}
&G_{p+q+2\alpha_2p,2\alpha_2p}^{\alpha_2p,p+q+\alpha_2p} \Bigg(\chi \left( \frac{1 + (\kappa-1)\beta}{\beta} \right)^{\frac{\alpha_2p}{r}} ~\bigg|~\begin{matrix} \kappa_3 \\ \kappa_4 \end{matrix} \Bigg) \underset{\mu_r \gg 1} \cong   \sum_{i=1}^{p+q+\alpha_2p}\left[ \chi \left( \frac{1 + (\kappa-1)\beta}{\beta} \right)^{\frac{\alpha_2p}{r}}\right]^{\kappa_{3,i}-1} \\&\times~\frac{\prod_{j=1,~j \neq i}^{p+q+\alpha_2p}\Gamma(\kappa_{3,i} - \kappa_{3,j})\prod_{j=1}^{\alpha_2p}\Gamma(1-\kappa_{3,i}+\kappa_{4,j})}{\prod_{j=\alpha_2p+1}^{2\alpha_2p}\Gamma(\kappa_{3,i}-\kappa_{4,j})\prod_{j=p+q+\alpha_2p+1}^{p+q+2\alpha_2p}\Gamma(\kappa_{3,j}-\kappa_{3,i}+1)}.
\end{split}
\end{equation}}

\textcolor{black}{From (\ref{eq:40}), the diversity gain achieved by the FSO bachauling system is $\min\left(\frac{\xi^2}{r},~\frac{m_1\alpha_1}{r},~\frac{m_2\alpha_2}{r}\right)$. Consequently, the diversity gain achieved by the hybrid system is obtained by
\begin{equation}
G_d = \min\left(\mathcal{G},~\min\left(\frac{\xi^2}{r},~\frac{m_1\alpha_1}{r},~\frac{m_2\alpha_2}{r}\right)\right).    
\end{equation}}
\subsection{Error Probability}
The probability of error averaged over the end-to-end SINDR realizations is given by
\begin{equation}
p_e(\delta,\tau,v,q_k) =\frac{\delta}{2\Gamma(\tau)}\sum_{k=1}^{v}\mathbb{E}_{\gamma_{e2e}}[\Gamma(\tau,q_k\gamma)],
\end{equation}

where $v,~\delta,~\tau$, and $q_k$ vary depending on the type of
detection (heterodyne technique or IM/DD) and modulation
being assumed. It is worth accentuating that this expression is
general enough to be used for both heterodyne and IM/DD
techniques and can be applicable to different modulation
schemes. The parameters $v,~\delta,~\tau$~, and $q_k$ are summarized in TABLE II. The derivation of the pdf of the SINDR is not tractable due to the presence of the hardware impairments parameters. Consequently, the probability of error will be evaluated based on the numerical integration.
\begin{table}[H]
\renewcommand{\arraystretch}{1.3}
\caption{\textsc{Parameters for Different Modulations$^\dag$}}
\label{tab:example}
\centering
\begin{tabular}{|c|c|c|c|c|c|}
    \hline
    \textbf{Modulation}  &  $\delta$ & $\tau$ & $q_k$ & $v$ & \textbf{Detection}\\
    \hline
   \textbf{OOK}    &  1& 0.5 & 0.5 & 1 & IM/DD\\
    \hline
   \textbf{BPSK}    &  1& 0.5 & 1 & 1 & Heterodyne\\
    \hline
     \textbf{M-PSK}    &  $\frac{2}{\max(\log_2(M), 2)}$ & 0.5 & $\text{sin}^2\left(\frac{(2k-1)\pi}{M}\right)$ & $\max(\frac{M}{4},1)$ & Heterodyne\\
    \hline
      \textbf{M-QAM}    & $\frac{4}{\log_2(M)}\left(1-\frac{1}{\sqrt{M}}  \right)$  & 0.5 & $\frac{3(2k-1)^2}{2(M-1)}$& $\frac{\sqrt{M}}{2}$& Heterodyne \\
    \hline
\end{tabular}
\\ 
\rule{0in}{1.2em}$^\dag$\scriptsize In case of OOK modulation, the parameters $v,~\delta,~\tau$~, and $q_k$ are given by \cite[Eq.~(26)]{hop1}. For M-PSK and M-QAM modulations, these parameters are provided by \cite[Eqs.~(30), (31)]{50}.
\end{table}
\subsection{Achievable Rate}
Given that the relay employs the DF scheme, the achievable rate of the hyrid system is given by
\begin{equation}
\mathcal{C} = \min(\mathcal{C}_1,\mathcal{C}_2).    
\end{equation}

 \textcolor{black}{An exact closed-form of the achievable rate is not tractable and hence a numerical integration is required. Note that the low and high SNR expansions along with the approximations and the upper bounds follow from the previous sections.}
 \textcolor{black}{\subsection{Rate Coverage}
 The rate coverage is defined as the probability when the achievable rate $\mathcal{C}$ is greater than a target rate, $r$, expressed in nats. It can be formulated as
 \begin{equation}
     \mathcal{R}_c(r) =\mathbb{P}[\mathcal{C}\geq r] = \mathbb{P}[\gamma_{e2e}\geq e^{\frac{r}{B}}-1].
 \end{equation}}
\section{Numerical Results and Discussion}
In this section, we validate the analylical expressions with the numerical numerical simulations using the Monte Carlo method \footnote{For all cases, $10^9$ realizations of the random variables were generated to perform the Monte Carlo simulation in MATLAB.}. Temporally correlated Gamma SNR coefficients are generated using (\ref{corr}). The atmospheric turbulence $I_a$ is generated using the expression $I_a = I_{x}\times I_{y}$, where the two independent random variables $I_{x}$ and $I_{y}$ follow the Generalized Gamma distribution using \cite{65}. In addition, the pointing errors is simulated by generating the radial displacement $R$ following the Rayleigh distribution with scale equal to the jitter standard deviation ($\sigma_s$) and then we generate the samples using (\ref{pointing}). Since the path loss is deterministic, it can be generated using the relation (\ref{pathloss}). TABLE \ref{fso-system} summarizes the main simulation parameters of the FSO sub-system.
\begin{table}[H]
\renewcommand{\arraystretch}{1}
\caption{FSO sub-system parameters}
\label{fso-system}
\centering
\begin{tabular}{|c|c|c|}
\hline
\bfseries Parameter & \bfseries Symbol & \bfseries Value\\
\hline
Wavelength & $\lambda_2$& 1550 nm\\
\hline
Receiver aperture radius & $a$& 5 cm\\
\hline
Divergence angle& $\theta$ & 10 mrad\\
\hline
Noise variance& $\sigma^2_{2}$& $\text{10}^{\text{-7}}$ A/Hz\\
\hline
Weather attenuation & $\sigma$ &  0.43 dB/km\\
\hline
Refractive index & $C^2_n$ & 5$\cdot$$\text{10}^{\text{-14}}$\\\hline
Link length & $L_2$ & 500 m\\\hline
\end{tabular}
\end{table}

Unless otherwise stated, the average number of the nearest BSs is $M=10$, the average number of interference is $M_z = 3$, the average receive power of the interference is 2 dB, the LOS and NLOS pathloss exponents are $\alpha_{\text{los}}=2$, and $\alpha_{\text{nlos}}=4$, respectively.  
Fig.~5.a shows the dependence of the outage performance with respect to different values of the target threshold $\beta$ and the correlation $\rho$. In addition, the relays are supposed to be impaired by the SEL impairments and the receiver employs the IM/DD as a method of detection. For both correlation values, we observe that the performance deteriorates as $\beta$ becomes higher and this result is certainly expected since for a given SNR, the probability that the SINDR falls below a higher threshold becomes higher. For a given threshold, the system works better when the best relay of the last rank ($k = M$) is selected according to PRS protocol. We observe that the performance improves as the correlation coefficient increases. For a perfect CSI estimation ($\rho = 0.9$), there are roughly a full correlation between the two CSIs and the selection of the best relay is certainly achieved based on the feedback or the outdated CSI. However, for a completely outdated CSI ($\rho = 0.1$) the two CSIs are completely uncorrelated and hence the selection of the best relay is uncertain since the selection is based on a completely outdated CSI. As a result, the performance deteriorates substantially.
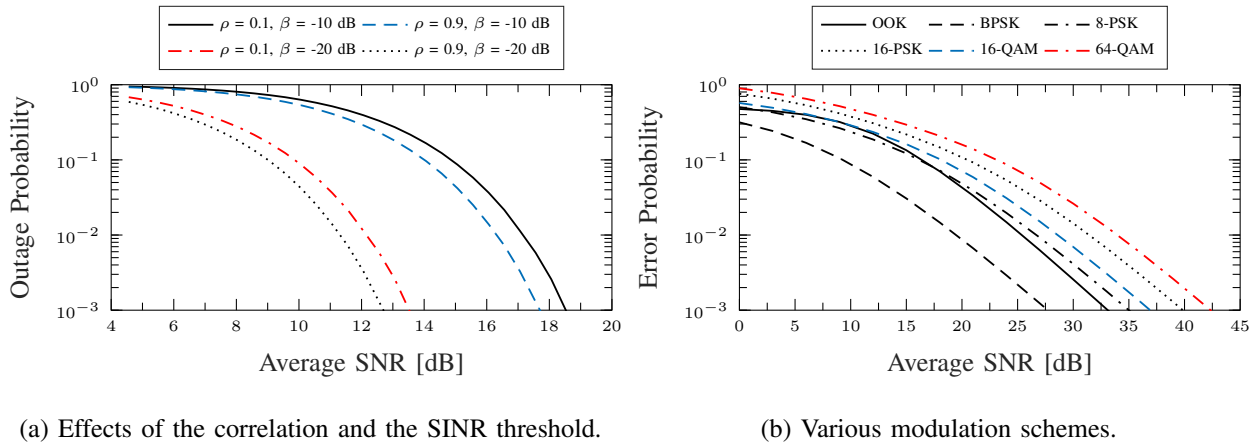
\begin{figure}[htbp]
\begin{subfigure}[b]{0.5\textwidth}
\centering
\setlength\fheight{3cm}
\setlength\fwidth{7cm}
%
%
\begin{tikzpicture}

\begin{axis}[%
width=0.951\fwidth,
height=\fheight,
at={(0\fwidth,0\fheight)},
scale only axis,
legend columns=2,
xmin=4,
xmax=20,
xlabel style={font=\color{white!15!black}},
xlabel={\small{Average SNR [dB]}},
ymode=log,
ymin=0.001,
ymax=1,
yminorticks=true,
ylabel style={font=\color{white!15!black}},
ylabel={\small{Outage Probability}},
axis background/.style={fill=white},
legend style={at={(0.5,1.35)}, anchor=north, legend cell align=left, align=left, draw=white!15!black}
]
\addplot [color=black, line width=0.7pt]
  table[row sep=crcr]{%
4.56197914243497	0.94658\\
5.06197914243497	0.93503\\
5.56197914243497	0.9213\\
6.06197914243497	0.90479\\
6.56197914243497	0.88497\\
7.06197914243497	0.86172\\
7.56197914243497	0.83473\\
8.06197914243497	0.80307\\
8.56197914243497	0.76703\\
9.06197914243497	0.7264\\
9.56197914243497	0.68158\\
10.061979142435	0.63016\\
10.561979142435	0.57413\\
11.061979142435	0.51614\\
11.561979142435	0.45504\\
12.061979142435	0.39241\\
12.561979142435	0.331\\
13.061979142435	0.27226\\
13.561979142435	0.21612\\
14.061979142435	0.16625\\
14.561979142435	0.12304\\
15.061979142435	0.08652\\
15.561979142435	0.05805\\
16.061979142435	0.03657\\
16.561979142435	0.02154\\
17.061979142435	0.01135\\
17.561979142435	0.00581\\
18.061979142435	0.00258\\
18.561979142435	0.00093\\
19.061979142435	0.00032\\
19.561979142435	0.00013\\
20.061979142435	6e-05\\
20.561979142435	0\\
21.061979142435	0\\
21.561979142435	0\\
22.061979142435	0\\
22.561979142435	0\\
23.061979142435	0\\
23.561979142435	0\\
24.061979142435	0\\
24.561979142435	0\\
25.061979142435	0\\
25.561979142435	0\\
26.061979142435	0\\
26.561979142435	0\\
27.061979142435	0\\
27.561979142435	0\\
28.061979142435	0\\
28.561979142435	0\\
29.061979142435	0\\
29.561979142435	0\\
30.061979142435	0\\
30.561979142435	0\\
31.061979142435	0\\
31.561979142435	0\\
32.061979142435	0\\
32.561979142435	0\\
33.061979142435	0\\
33.561979142435	0\\
34.061979142435	0\\
34.561979142435	0\\
};
\addlegendentry{\tiny{$\rho\text{ = 0.1, }\beta\text{ = -10 dB}$}}

\addplot [color=NavyBlue, dash pattern={on 5pt off 3pt on 0pt off 0pt} , line width=0.7pt]
  table[row sep=crcr]{%
4.56197914243497	0.92526\\
5.06197914243497	0.90966\\
5.56197914243497	0.89103\\
6.06197914243497	0.86866\\
6.56197914243497	0.84277\\
7.06197914243497	0.81252\\
7.56197914243497	0.77776\\
8.06197914243497	0.73874\\
8.56197914243497	0.6948\\
9.06197914243497	0.64512\\
9.56197914243497	0.59074\\
10.061979142435	0.53327\\
10.561979142435	0.47332\\
11.061979142435	0.4098\\
11.561979142435	0.34841\\
12.061979142435	0.28833\\
12.561979142435	0.23188\\
13.061979142435	0.17982\\
13.561979142435	0.13425\\
14.061979142435	0.09632\\
14.561979142435	0.06577\\
15.061979142435	0.04196\\
15.561979142435	0.0251\\
16.061979142435	0.0139\\
16.561979142435	0.00725\\
17.061979142435	0.00329\\
17.561979142435	0.00129\\
18.061979142435	0.00051\\
18.561979142435	0.00017\\
19.061979142435	6e-05\\
19.561979142435	1e-05\\
20.061979142435	0\\
20.561979142435	0\\
21.061979142435	0\\
21.561979142435	0\\
22.061979142435	0\\
22.561979142435	0\\
23.061979142435	0\\
23.561979142435	0\\
24.061979142435	0\\
24.561979142435	0\\
25.061979142435	0\\
25.561979142435	0\\
26.061979142435	0\\
26.561979142435	0\\
27.061979142435	0\\
27.561979142435	0\\
28.061979142435	0\\
28.561979142435	0\\
29.061979142435	0\\
29.561979142435	0\\
30.061979142435	0\\
30.561979142435	0\\
31.061979142435	0\\
31.561979142435	0\\
32.061979142435	0\\
32.561979142435	0\\
33.061979142435	0\\
33.561979142435	0\\
34.061979142435	0\\
34.561979142435	0\\
};
\addlegendentry{\tiny{$\rho\text{ = 0.9, }\beta\text{ = -10 dB}$}}

\addplot [color=red, dash pattern={on 5pt off 3pt on 1pt off 3pt} , line width=0.7pt]
  table[row sep=crcr]{%
4.56197914243497	0.68155\\
5.06197914243497	0.6301\\
5.56197914243497	0.57418\\
6.06197914243497	0.51636\\
6.56197914243497	0.45488\\
7.06197914243497	0.39224\\
7.56197914243497	0.331\\
8.06197914243497	0.27227\\
8.56197914243497	0.21609\\
9.06197914243497	0.16645\\
9.56197914243497	0.12318\\
10.061979142435	0.08644\\
10.561979142435	0.05797\\
11.061979142435	0.03639\\
11.561979142435	0.02154\\
12.061979142435	0.0113\\
12.561979142435	0.00583\\
13.061979142435	0.00259\\
13.561979142435	0.00092\\
14.061979142435	0.00032\\
14.561979142435	0.00015\\
15.061979142435	6e-05\\
15.561979142435	0\\
16.061979142435	0\\
16.561979142435	0\\
17.061979142435	0\\
17.561979142435	0\\
18.061979142435	0\\
18.561979142435	0\\
19.061979142435	0\\
19.561979142435	0\\
20.061979142435	0\\
20.561979142435	0\\
21.061979142435	0\\
21.561979142435	0\\
22.061979142435	0\\
22.561979142435	0\\
23.061979142435	0\\
23.561979142435	0\\
24.061979142435	0\\
24.561979142435	0\\
25.061979142435	0\\
25.561979142435	0\\
26.061979142435	0\\
26.561979142435	0\\
27.061979142435	0\\
27.561979142435	0\\
28.061979142435	0\\
28.561979142435	0\\
29.061979142435	0\\
29.561979142435	0\\
30.061979142435	0\\
30.561979142435	0\\
31.061979142435	0\\
31.561979142435	0\\
32.061979142435	0\\
32.561979142435	0\\
33.061979142435	0\\
33.561979142435	0\\
34.061979142435	0\\
34.561979142435	0\\
};
\addlegendentry{\tiny{$\rho\text{ = 0.1, }\beta\text{ = -20 dB}$}}

\addplot [color=black, dotted, line width=0.7pt]
  table[row sep=crcr]{%
4.56197914243497	0.5906\\
5.06197914243497	0.53364\\
5.56197914243497	0.47339\\
6.06197914243497	0.40992\\
6.56197914243497	0.34846\\
7.06197914243497	0.2884\\
7.56197914243497	0.23167\\
8.06197914243497	0.17978\\
8.56197914243497	0.13434\\
9.06197914243497	0.09646\\
9.56197914243497	0.06574\\
10.061979142435	0.04203\\
10.561979142435	0.02507\\
11.061979142435	0.01389\\
11.561979142435	0.00719\\
12.061979142435	0.00327\\
12.561979142435	0.00129\\
13.061979142435	0.00051\\
13.561979142435	0.00018\\
14.061979142435	6e-05\\
14.561979142435	1e-05\\
15.061979142435	0\\
15.561979142435	0\\
16.061979142435	0\\
16.561979142435	0\\
17.061979142435	0\\
17.561979142435	0\\
18.061979142435	0\\
18.561979142435	0\\
19.061979142435	0\\
19.561979142435	0\\
20.061979142435	0\\
20.561979142435	0\\
21.061979142435	0\\
21.561979142435	0\\
22.061979142435	0\\
22.561979142435	0\\
23.061979142435	0\\
23.561979142435	0\\
24.061979142435	0\\
24.561979142435	0\\
25.061979142435	0\\
25.561979142435	0\\
26.061979142435	0\\
26.561979142435	0\\
27.061979142435	0\\
27.561979142435	0\\
28.061979142435	0\\
28.561979142435	0\\
29.061979142435	0\\
29.561979142435	0\\
30.061979142435	0\\
30.561979142435	0\\
31.061979142435	0\\
31.561979142435	0\\
32.061979142435	0\\
32.561979142435	0\\
33.061979142435	0\\
33.561979142435	0\\
34.061979142435	0\\
34.561979142435	0\\
};
\addlegendentry{\tiny{$\rho\text{ = 0.9, }\beta\text{ = -20 dB}$}}

\end{axis}
\end{tikzpicture}%
    \caption{Effects of the correlation and the SINR threshold.}
    \label{a1}
    \end{subfigure}
    \begin{subfigure}[b]{0.5\textwidth}
\centering
\setlength\fheight{3cm}
\setlength\fwidth{7cm}
%
%
\begin{tikzpicture}

\begin{axis}[%
width=0.951\fwidth,
height=\fheight,
at={(0\fwidth,0\fheight)},
scale only axis,
legend columns=3,
xmin=0,
xmax=45,
xlabel style={font=\color{white!15!black}},
xlabel={\small{Average SNR [dB]}},
xtick={0,5,10,15,20,25,30,35,40,45},
ymode=log,
ymin=1e-03,
ymax=1,
yminorticks=true,
ylabel style={font=\color{white!15!black}},
ylabel={\small{Error Probability}},
axis background/.style={fill=white},
legend style={at={(0.5,1.35)}, anchor=north, legend cell align=left, align=left, draw=white!15!black}
]
\addplot [color=black, line width=0.7pt]
  table[row sep=crcr]{%
0	0.4756104640225\\
3	0.448140408963638\\
6	0.397928318902997\\
9	0.319652055136726\\
12	0.223535475813997\\
15	0.134283117990849\\
18	0.0704736247077663\\
21	0.0334295274496445\\
24	0.0148104916290612\\
27	0.00626653215745328\\
30	0.00256408007964772\\
33	0.00103241963190347\\
36	0.000411771034003602\\
39	0.000159529556142667\\
42	5.90429444708341e-05\\
45	2.00362133644383e-05\\
};
\addlegendentry{\tiny{OOK}}

\addplot [color=black, dash pattern={on 5pt off 3pt on 0pt off 0pt} , line width=0.7pt]
  table[row sep=crcr]{%
0	0.315818761471881\\
3	0.242139272556022\\
6	0.168417943834438\\
9	0.105398232066377\\
12	0.0594939680631021\\
15	0.0306700438119262\\
18	0.0147009922157151\\
21	0.00667114719262079\\
24	0.00290820241463642\\
27	0.00123193116867219\\
30	0.000512517656493201\\
33	0.000210119442264207\\
36	8.37605937914893e-05\\
39	3.19117666553922e-05\\
42	1.16261661601415e-05\\
45	4.23143971002138e-06\\
};
\addlegendentry{\tiny{BPSK}}

\addplot [color=black, dash pattern={on 5pt off 3pt on 1pt off 3pt} , line width=0.7pt]
  table[row sep=crcr]{%
0	0.501753413129372\\
3	0.429137450581642\\
6	0.347622747788907\\
9	0.264430896145889\\
12	0.187076429241043\\
15	0.121770211495161\\
18	0.0723608256308218\\
21	0.0392778906634092\\
24	0.0196701214640476\\
27	0.00923105827242278\\
30	0.00412593538709129\\
33	0.00177924150850974\\
36	0.000747258266055834\\
39	0.000308023158380688\\
42	0.00012466523376561\\
45	4.89283774132807e-05\\
};
\addlegendentry{\tiny{8-PSK}}

\addplot [color=black, dotted, line width=0.7pt]
  table[row sep=crcr]{%
0	0.757714734600896\\
3	0.651635239773983\\
6	0.533671372922961\\
9	0.415435091869847\\
12	0.308067473994296\\
15	0.218050357450168\\
18	0.146576490119473\\
21	0.0923460330217237\\
24	0.0537931250966397\\
27	0.0288406883350517\\
30	0.014329930584759\\
33	0.00669095832172965\\
36	0.00297925459717401\\
39	0.00127940953585111\\
42	0.000534810938341631\\
45	0.000219687518492584\\
};
\addlegendentry{\tiny{16-PSK}}

\addplot [color=NavyBlue, dash pattern={on 5pt off 3pt on 0pt off 0pt} , line width=0.7pt]
  table[row sep=crcr]{%
0	0.571094138833353\\
3	0.493230128933118\\
6	0.406598754781537\\
9	0.318468149502236\\
12	0.235289040537518\\
15	0.161907120671499\\
18	0.102287049502803\\
21	0.0588749507578893\\
24	0.0309915422193991\\
27	0.0151192579878143\\
30	0.0069532227459382\\
33	0.00306092326952077\\
36	0.00130419987484732\\
39	0.000543156805531503\\
42	0.000222699886953452\\
45	8.94104959798147e-05\\
};
\addlegendentry{\tiny{16-QAM}}

\addplot [color=red, dash pattern={on 5pt off 3pt on 1pt off 3pt} , line width=0.7pt]
  table[row sep=crcr]{%
0	0.895748675743705\\
3	0.778575036747088\\
6	0.648909392607065\\
9	0.518404747335482\\
12	0.397844807882967\\
15	0.293851321173178\\
18	0.208101600781783\\
21	0.139497550738815\\
24	0.0869628637783266\\
27	0.0498026034097663\\
30	0.0261961803512389\\
33	0.0127912809495045\\
36	0.00588852102545877\\
39	0.00259323011429146\\
42	0.00110453598539519\\
45	0.000459618196470623\\
};
\addlegendentry{\tiny{64-QAM}}

\end{axis}
\end{tikzpicture}%
    \caption{Various modulation schemes.}
    \label{b1}
    \end{subfigure}
    \caption[map]{System performance. (a) Probability of outage where the impairment model is SEL and IM/DD is the detection technique. (b) Probability of error for different modulation schemes considering ideal hardware. }
    \label{rfidtag_testing}
\end{figure}
Fig.~5.b illustrates the variations of the probability of error for various modulation schemes. We observe that the system works better for BPSK, however, the performance gets much worse for 64-QAM modulation. In fact, there is a tradeoff between these two modulation schemes: BPSK yields lower error while the 64-QAM provides much more bandwidth efficiency which is very advantageous. For practice uses, mmWaves system cannot exceed the QPSK constellation as the average SNR is very low and the error will be significant resulting in a low achievable rate.

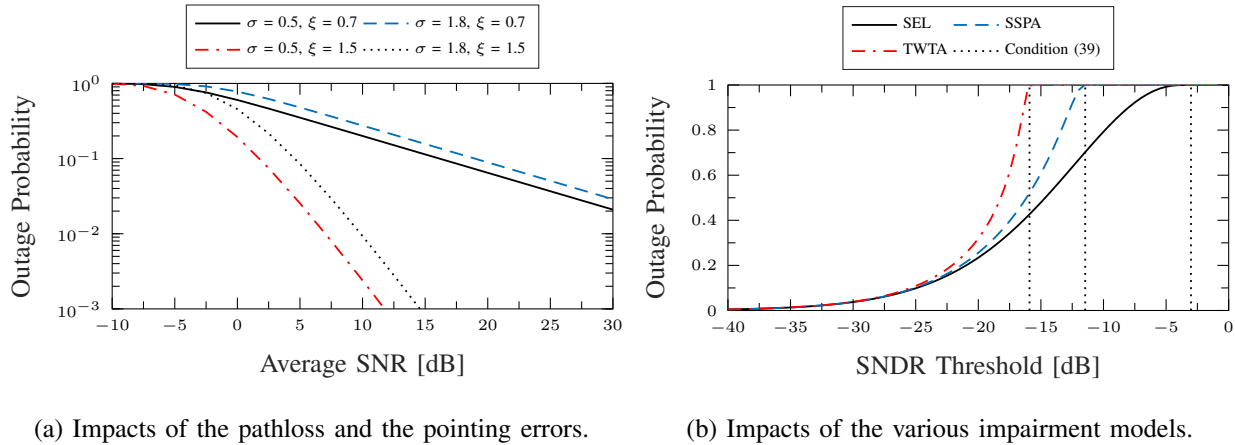
\begin{figure}[htbp]
\begin{subfigure}[b]{0.5\textwidth}
\centering
\setlength\fheight{3cm}
\setlength\fwidth{7cm}
%
%
\begin{tikzpicture}

\begin{axis}[%
width=0.951\fwidth,
height=\fheight,
at={(0\fwidth,0\fheight)},
scale only axis,
xmin=-10,
xmax=30,
xtick = {-10,-5,0,5,10,15,20,25,30},
legend columns=2,
xlabel style={font=\color{white!15!black}},
xlabel={\small{Average SNR [dB]}},
ymode=log,
ymin=0.001,
ymax=1,
yminorticks=true,
ylabel style={font=\color{white!15!black}},
ylabel={\small{Outage Probability}},
axis background/.style={fill=white},
legend style={at={(0.5,1.35)}, anchor=north, legend cell align=left, align=left, draw=white!15!black}
]
\addplot [color=black, line width=0.7pt]
  table[row sep=crcr]{%
-10	0.998153317851169\\
-7.5	0.977071782204861\\
-5	0.898427614378967\\
-2.5	0.75917044729216\\
0	0.602004466709541\\
2.5	0.462251639059315\\
5	0.350561169578634\\
7.5	0.264803194563444\\
10	0.199799286220989\\
12.5	0.150708036709119\\
15	0.113670230544671\\
17.5	0.0857332364730119\\
20	0.0646621006012792\\
22.5	0.0487696848075444\\
25	0.0367832400922765\\
27.5	0.0277427807518802\\
30	0.0209242543977985\\
};
\addlegendentry{\tiny{$\sigma\text{ = 0.5, }\xi\text{ = 0.7}$}}

\addplot [color=NavyBlue, dash pattern={on 5pt off 3pt on 0pt off 0pt} , line width=0.7pt]
  table[row sep=crcr]{%
-10	0.999985369322141\\
-7.5	0.998797157357881\\
-5	0.982266946024109\\
-2.5	0.912582527164061\\
0	0.779193494957813\\
2.5	0.6217998286348\\
5	0.478764624221446\\
7.5	0.363429166508907\\
10	0.274600756813927\\
12.5	0.20720753389604\\
15	0.156299094687196\\
17.5	0.117887802480068\\
20	0.0889143490869472\\
22.5	0.0670613920060672\\
25	0.0505792904334147\\
27.5	0.0381480876917698\\
30	0.0287721808678259\\
};
\addlegendentry{\tiny{$\sigma\text{ = 1.8, }\xi\text{ = 0.7}$}}

\addplot [color=red,dash pattern={on 5pt off 3pt on 1pt off 3pt}, line width=0.7pt]
  table[row sep=crcr]{%
-10	0.993293336266327\\
-7.5	0.925034396402675\\
-5	0.710746448641867\\
-2.5	0.418565140278271\\
0	0.193395155359822\\
2.5	0.0745506139722618\\
5	0.0254111974096089\\
7.5	0.00799688245012732\\
10	0.00239198057448384\\
12.5	0.000692875097114448\\
15	0.000196670472067231\\
17.5	5.51101029410183e-05\\
20	1.5316470346133e-05\\
22.5	4.23451853256076e-06\\
25	1.16676361697491e-06\\
27.5	3.20786857007654e-07\\
30	8.80723849740935e-08\\
};
\addlegendentry{\tiny{$\sigma\text{ = 0.5, }\xi\text{ = 1.5}$}}

\addplot [color=black, dotted, line width=0.7pt]
  table[row sep=crcr]{%
-10	0.999942276684855\\
-7.5	0.995582171040034\\
-5	0.941118645573583\\
-2.5	0.746127537522564\\
0	0.454866031303309\\
2.5	0.216159683347666\\
5	0.0849972830275906\\
7.5	0.0293549224709511\\
10	0.00931604004770853\\
12.5	0.00280144558657947\\
15	0.000814218416345881\\
17.5	0.000231606316011784\\
20	6.49876472121205e-05\\
22.5	1.80773233541657e-05\\
25	5.00058188421658e-06\\
27.5	1.3783341999329e-06\\
30	3.79042867223026e-07\\
};
\addlegendentry{\tiny{$\sigma\text{ = 1.8, }\xi\text{ = 1.5}$}}

\end{axis}
\end{tikzpicture}%
    \caption{Impacts of the pathloss and the pointing errors.}
    \label{a1}
    \end{subfigure}
    \begin{subfigure}[b]{0.5\textwidth}
\centering
\setlength\fheight{3cm}
\setlength\fwidth{7cm}
%
%
\begin{tikzpicture}

\begin{axis}[%
width=0.951\fwidth,
height=\fheight,
at={(0\fwidth,0\fheight)},
scale only axis,
xmin=-40,
xmax=0,
legend columns=2,
xlabel style={font=\color{white!15!black}},
xlabel={\small{SNDR Threshold [dB]}},
xtick={-40,-35,-30,-25,-20,-15,-10,-5,0},
ymin=0,
ymax=1,
ylabel style={font=\color{white!15!black}},
ylabel={\small{Outage Probability}},
axis background/.style={fill=white},
legend style={at={(0.5,1.35)}, anchor=north, legend cell align=left, align=left, draw=white!15!black}
]
\addplot [color=black, line width=0.7pt]
  table[row sep=crcr]{%
-40	0.00464735802402678\\
-39.5	0.00517369137652017\\
-39	0.00575843475534671\\
-38.5	0.00640785298343906\\
-38	0.00712884375143365\\
-37.5	0.00792899454720043\\
-37	0.00881664344010787\\
-36.5	0.00980094374001334\\
-36	0.0108919324940773\\
-35.5	0.0121006027144967\\
-35	0.0134389791453212\\
-34.5	0.0149201972747472\\
-34	0.0165585851786787\\
-33.5	0.0183697476399085\\
-33	0.0203706518230144\\
-32.5	0.0225797135961321\\
-32	0.025016883375494\\
-31.5	0.0277037301257054\\
-31	0.0306635218773066\\
-30.5	0.0339213008230782\\
-30	0.037503950726441\\
-29.5	0.0414402540209666\\
-29	0.0457609356025565\\
-28.5	0.0504986899200232\\
-28	0.0556881875623013\\
-27.5	0.0613660571302395\\
-27	0.0675708377793343\\
-26.5	0.0743428974410802\\
-26	0.081724311392089\\
-25.5	0.0897586955621449\\
-25	0.098490988778421\\
-24.5	0.10796717805974\\
-24	0.11823396113116\\
-23.5	0.129338340556581\\
-23	0.141327144317881\\
-22.5	0.154246468335737\\
-22	0.168141037360501\\
-21.5	0.183053481888607\\
-21	0.199023530302397\\
-20.5	0.216087117302022\\
-20	0.23427541189898\\
-19.5	0.253613770759581\\
-19	0.274120625493628\\
-18.5	0.295806315530075\\
-18	0.318671881436602\\
-17.5	0.342707836830862\\
-17	0.367892940282021\\
-16.5	0.394192991675915\\
-16	0.421559680263187\\
-15.5	0.44992951386414\\
-15	0.479222860303271\\
-14.5	0.509343132938464\\
-14	0.540176152010837\\
-13.5	0.571589712396688\\
-13	0.603433386194348\\
-12.5	0.635538585539892\\
-12	0.667718907391425\\
-11.5	0.699770778260569\\
-11	0.731474413854796\\
-10.5	0.762595107705319\\
-10	0.792884866309289\\
-9.5	0.822084419713647\\
-9	0.849925661789143\\
-8.5	0.876134623880227\\
-8	0.900435176963736\\
-7.5	0.922553823243048\\
-7	0.942226238306998\\
-6.5	0.95920677385622\\
-6	0.973283149348302\\
-5.5	0.984300472881434\\
-5	0.992202293690752\\
-4.5	0.997102488750923\\
-4	0.999407799048377\\
-3.5	0.999979467185691\\
-3	1\\
-2.5	1\\
-2	1\\
-1.5	1\\
-1	1\\
-0.5	1\\
0	1\\
0.5	1\\
1	1\\
1.5	1\\
2	1\\
2.5	1\\
3	1\\
3.5	1\\
4	1\\
4.5	1\\
5	1\\
5.5	1\\
6	1\\
6.5	1\\
7	1\\
7.5	1\\
8	1\\
8.5	1\\
9	1\\
9.5	1\\
10	1\\
};
\addlegendentry{\tiny{SEL}}

\addplot [color=NavyBlue, dash pattern={on 5pt off 3pt on 0pt off 0pt} , line width=0.7pt]
  table[row sep=crcr]{%
-40	0.00465256584517251\\
-39.5	0.00518018506967832\\
-39	0.00576652932304548\\
-38.5	0.00641793980110691\\
-38	0.00714140883381992\\
-37.5	0.00794464110559872\\
-37	0.00883611971428248\\
-36.5	0.00982517730707211\\
-36	0.0109220725211715\\
-35.5	0.0121380719423499\\
-35	0.013485537776033\\
-34.5	0.0149780214028111\\
-34	0.0166303629636012\\
-33.5	0.0184587970895674\\
-33	0.020481064859092\\
-32.5	0.0227165320298424\\
-32	0.0251863135601015\\
-31.5	0.0279134044025263\\
-31	0.0309228165286879\\
-30.5	0.0342417221284851\\
-30	0.0378996029303449\\
-29.5	0.0419284056129956\\
-29	0.0463627033361301\\
-28.5	0.0512398635160428\\
-28	0.056600222126088\\
-27.5	0.0624872650259743\\
-27	0.0689478171368217\\
-26.5	0.0760322407023706\\
-26	0.0837946444365192\\
-25.5	0.0922931060838256\\
-25	0.101589911848423\\
-24.5	0.111751817319718\\
-24	0.122850335989003\\
-23.5	0.134962063266045\\
-23	0.148169046133135\\
-22.5	0.162559211287131\\
-22	0.178226867892181\\
-21.5	0.195273304965219\\
-21	0.213807507983493\\
-20.5	0.233947024505791\\
-20	0.255819014237541\\
-19.5	0.279561524487605\\
-19	0.305325036056896\\
-18.5	0.333274324411533\\
-18	0.363590670446285\\
-17.5	0.396474421357559\\
-17	0.432147817602149\\
-16.5	0.470857806702339\\
-16	0.512878124547562\\
-15.5	0.558508927600914\\
-15	0.608069943133885\\
-14.5	0.661877496180468\\
-14	0.7201813712433\\
-13.5	0.782997475985725\\
-13	0.849649233541867\\
-12.5	0.917399423080349\\
-12	0.976835215446866\\
-11.5	0.999999530332645\\
-11	1\\
-10.5	1\\
-10	1\\
-9.5	1\\
-9	1\\
-8.5	1\\
-8	1\\
-7.5	1\\
-7	1\\
-6.5	1\\
-6	1\\
-5.5	1\\
-5	1\\
-4.5	1\\
-4	1\\
-3.5	1\\
-3	1\\
-2.5	1\\
-2	1\\
-1.5	1\\
-1	1\\
-0.5	1\\
0	1\\
0.5	1\\
1	1\\
1.5	1\\
2	1\\
2.5	1\\
3	1\\
3.5	1\\
4	1\\
4.5	1\\
5	1\\
5.5	1\\
6	1\\
6.5	1\\
7	1\\
7.5	1\\
8	1\\
8.5	1\\
9	1\\
9.5	1\\
10	1\\
};
\addlegendentry{\tiny{SSPA}}

\addplot [color=red, dash pattern={on 5pt off 3pt on 1pt off 3pt}, line width=0.7pt]
  table[row sep=crcr]{%
-40	0.00466345408632336\\
-39.5	0.00519376756630233\\
-39	0.00578346841087367\\
-38.5	0.00643905932179191\\
-38	0.00716773312837235\\
-37.5	0.00797744331206784\\
-37	0.00887698156039855\\
-36.5	0.00987606309737265\\
-36	0.0109854206438238\\
-35.5	0.012216907990271\\
-35	0.0135836143230924\\
-34.5	0.0150999906398301\\
-34	0.0167819898313324\\
-33.5	0.0186472223101002\\
-33	0.0207151294422207\\
-32.5	0.0230071775161098\\
-32	0.0255470755827853\\
-31.5	0.0283610212658533\\
-31	0.0314779796124347\\
-30.5	0.0349300013017386\\
-30	0.0387525881295024\\
-29.5	0.0429851157561454\\
-29	0.0476713263958498\\
-28.5	0.0528599076400046\\
-28	0.0586051782374214\\
-27.5	0.0649679077944758\\
-27	0.0720163055744766\\
-26.5	0.0798272246733371\\
-26	0.0884876429968665\\
-25.5	0.0980965033827679\\
-25	0.108767024464866\\
-24.5	0.120629635365229\\
-24	0.133835747042588\\
-23.5	0.148562660603369\\
-23	0.165020043333237\\
-22.5	0.183458601676946\\
-22	0.204181889052618\\
-21.5	0.227562678123868\\
-21	0.254066131572227\\
-20.5	0.284283359978312\\
-20	0.318981308884655\\
-19.5	0.359179142623085\\
-19	0.406269123786934\\
-18.5	0.462214828416012\\
-18	0.529887194377553\\
-17.5	0.613640585012595\\
-17	0.720184324504775\\
-16.5	0.858130381831786\\
-16	0.997184609590086\\
-15.5	1\\
-15	1\\
-14.5	1\\
-14	1\\
-13.5	1\\
-13	1\\
-12.5	1\\
-12	1\\
-11.5	1\\
-11	1\\
-10.5	1\\
-10	1\\
-9.5	1\\
-9	1\\
-8.5	1\\
-8	1\\
-7.5	1\\
-7	1\\
-6.5	1\\
-6	1\\
-5.5	1\\
-5	1\\
-4.5	1\\
-4	1\\
-3.5	1\\
-3	1\\
-2.5	1\\
-2	1\\
-1.5	1\\
-1	1\\
-0.5	1\\
0	1\\
0.5	1\\
1	1\\
1.5	1\\
2	1\\
2.5	1\\
3	1\\
3.5	1\\
4	1\\
4.5	1\\
5	1\\
5.5	1\\
6	1\\
6.5	1\\
7	1\\
7.5	1\\
8	1\\
8.5	1\\
9	1\\
9.5	1\\
10	1\\
};
\addlegendentry{\tiny{TWTA}}

\addplot [color=black, dotted, line width=0.7pt]
  table[row sep=crcr]{%
-3.01029995663981	0\\
-3.01029995663981	1\\
};
\addlegendentry{\tiny{Condition (\ref{cdf-sndr})}}

\addplot [color=black, dotted, line width=0.7pt, forget plot]
  table[row sep=crcr]{%
-11.4612803567824	0\\
-11.4612803567824	1\\
};
\addplot [color=black, dotted, line width=0.7pt, forget plot]
  table[row sep=crcr]{%
-15.910646070265	0\\
-15.910646070265	1\\
};
\end{axis}
\end{tikzpicture}%
    \caption{Impacts of the various impairment models.}
    \label{b1}
    \end{subfigure}
    \caption[map]{Outage performance. (a) The impairment model is SSPA while the heterodyne detection is assumed. (b) The outage is evaluated with respect to the SNDR threshold accounting for the necessary condition. }
    \label{rfidtag_testing}
\end{figure}

Fig.~6.a illustrates the impacts of the FSO atmoshperic pathloss and the pointing errors. Basically, the pathloss incurs a performance loss to the system and this loss gap increases with the severity of the pathloss. Most importantly, the system depends to a large extent on the pointing errors severity. The losses are mainly expressed as a significant reduction in the diversity gain of the system.

Fig.~6.b presents the dependence of the outage performance on the impairment models. We observe that the probability of outage saturates at different SNDR thresholds constrained by the necessary condition. Effectively, each condition depends on the parameters of the impairment models. We note that TWTA is more severe compared to SSPA and SEL as the outage saturates at relatively low SNDR threshold around -14 dB, while the system saturates at roughly -11 dB, and -3 dB for SSPA, and SEL, respectively.

\textcolor{black}{Fig.~7.a illustrates a different way to interpret the losses created by the hardware impairments. The probability of outage is evaluated with respect to the average SNR for the different impairment models. In this simulation, the loss severity is measured with respect to the level of the outage floor. The higher the outage floor is, the higher the losses are. In an agreement with the conluding remarks drawn for Fig. 6.b, the TWTA introduces an irreducible high floor level compared to SSPA and SEL which exhibit lower outage floors.} 

\textcolor{black}{Fig.~7.b presents a comparison in terms of the achievable rates for mmWaves and sub-6 GHz carrier frequencies. For sub-6 GHz configuration, we set 4 transmit antennas at the BS and we assume a bandwidth of 10 MHz. At lower distances up to 400 meters, mmWave achieves higher rate compared to sub-6 GHz, however, mmWave performance severely degrades at higher distances resulting in low achievable rate compared to sub-6 GHz. This result is explained by the fact that higher frequencies are significantly attenuated by the pathloss. In this simulation, the array gain efficiently compensates for the pathloss up to 400 meters, however, such gain becomes insufficient to compensate for the pathloss which becomes more severe for longer distances. Thereby, these observations confirm that mmWaves are more suitable for small densified cells where high data rate is required, while sub-6 GHz is more relevant for large cells where an acceptable rate coverage is still achieved.}
\begin{figure}[htbp]
\begin{subfigure}[b]{0.5\textwidth}
\centering
\setlength\fheight{3cm}
\setlength\fwidth{7cm}
%
%
\begin{tikzpicture}

\begin{axis}[%
width=0.951\fwidth,
height=\fheight,
at={(0\fwidth,0\fheight)},
scale only axis,
xmin=-10,
xmax=40,
xlabel style={font=\color{white!15!black}},
xlabel={\small{Average SNR [dB]}},
ymode=log,
ymin=0.001,
ymax=1,
yminorticks=true,
xtick={-10,-5,0,5,10,15,20,25,30,35,40},
ylabel style={font=\color{white!15!black}},
ylabel={\small{Outage Probability}},
axis background/.style={fill=white},
legend style={at={(0.03,0.03)}, anchor=south west, legend cell align=left, align=left, draw=white!15!black}
]
\addplot [color=black, line width=0.7pt]
  table[row sep=crcr]{%
-10	1\\
-8	0.999999999999999\\
-6	0.99999999896862\\
-4	0.999995234412669\\
-2	0.999097851485731\\
0	0.977780463762256\\
2	0.854870306207195\\
4	0.599102616315402\\
6	0.340162621195683\\
8	0.17355433941315\\
10	0.0879170897782737\\
12	0.0464991122802001\\
14	0.0258957781810173\\
16	0.015038350221741\\
18	0.00899319569582935\\
20	0.00548305906112012\\
22	0.00338476707603375\\
24	0.00210608276370772\\
26	0.00131707959895655\\
28	0.000826310963951693\\
30	0.000519475155163041\\
32	0.000327005007009551\\
34	0.000206019093644816\\
36	0.000129865306243926\\
38	8.18894125473868e-05\\
40	5.16485062376226e-05\\
};
\addlegendentry{\tiny{Ideal}}

\addplot [color=NavyBlue, dash pattern={on 5pt off 3pt on 0pt off 0pt} , line width=0.7pt]
  table[row sep=crcr]{%
-10	1\\
-8	0.999999999999999\\
-6	0.999999999221548\\
-4	0.999996110496247\\
-2	0.999225666228085\\
0	0.980259122522875\\
2	0.86742504154497\\
4	0.624345270220721\\
6	0.368003584594131\\
8	0.195508376865204\\
10	0.102926698847281\\
12	0.0565336020677935\\
14	0.0328821392152736\\
16	0.0202334120436006\\
18	0.0131332036586707\\
20	0.00899166051905764\\
22	0.00650892025122662\\
24	0.00499277471722337\\
26	0.00405552936350517\\
28	0.0034715254228902\\
30	0.00310575666979429\\
32	0.0028759139193143\\
34	0.00273117850666738\\
36	0.00263991270889641\\
38	0.00258231326100622\\
40	0.00254594106747141\\
};
\addlegendentry{\tiny{SEL}}

\addplot [color=red, dash pattern={on 5pt off 3pt on 1pt off 3pt} , line width=0.7pt]
  table[row sep=crcr]{%
-10	1\\
-8	0.999999999999999\\
-6	0.999999999295609\\
-4	0.999996381222481\\
-2	0.999266534568294\\
0	0.981070702240907\\
2	0.871627926310175\\
4	0.633019395102164\\
6	0.377871528606789\\
8	0.203564350201384\\
10	0.108640813516477\\
12	0.0604984136280859\\
14	0.0357430778600127\\
16	0.0224309466937865\\
18	0.0149332978994211\\
20	0.0105508315431794\\
22	0.00791997190906579\\
24	0.00631165610633988\\
26	0.0053165302194198\\
28	0.00469594297262088\\
30	0.00430694958765254\\
32	0.00406231967744064\\
34	0.00390815144514478\\
36	0.0038108612515777\\
38	0.00374941149265084\\
40	0.00371057749934445\\
};
\addlegendentry{\tiny{SSPA}}

\addplot [color=black, dotted, line width=0.7pt]
  table[row sep=crcr]{%
-10	1\\
-8	1\\
-6	0.999999999752207\\
-4	0.999998295448525\\
-2	0.999583040748099\\
0	0.987775002810946\\
2	0.908461954171452\\
4	0.714424632528069\\
6	0.478221277594049\\
8	0.293047184374574\\
10	0.178128007102032\\
12	0.113099034894352\\
14	0.0768003121176259\\
16	0.0561344669701554\\
18	0.0440359557617921\\
20	0.0367719516638735\\
22	0.032324533803018\\
24	0.029563149773125\\
26	0.0278320239650439\\
28	0.0267397541173465\\
30	0.0260476516425604\\
32	0.0256079070672728\\
34	0.0253280167668469\\
36	0.0251496772321192\\
38	0.0250359685713412\\
40	0.024963440778065\\
};
\addlegendentry{\tiny{TWTA}}

\end{axis}
\end{tikzpicture}%
    \caption{Illustration of the outage floors.}
    \label{a1}
    \end{subfigure}
    \begin{subfigure}[b]{0.5\textwidth}
\centering
\setlength\fheight{3cm}
\setlength\fwidth{7cm}
%
%
\begin{tikzpicture}

\begin{axis}[%
width=0.951\fwidth,
height=\fheight,
at={(0\fwidth,0\fheight)},
scale only axis,
xmin=0,
xmax=1000,
xlabel style={font=\color{white!15!black}},
xlabel={\small{Distance [m]}},
ymode=log,
ymin=10,
ymax=1000000000,
xtick={0,100,200,300,400,500,600,700,800,900,1000},
ytick = {10,100,1e3,1e4,1e5,1e6,1e7,1e8,1e9},
yminorticks=true,
ylabel style={font=\color{white!15!black}},
ylabel={\small{Achievable Rate [nats]}},
axis background/.style={fill=white},
legend style={legend cell align=left, align=left, draw=white!15!black}
]
\addplot [color=black, line width=0.7pt]
  table[row sep=crcr]{%
50	369557110.488618\\
100	76500073.9020161\\
150	17918223.7257716\\
200	4765342.27995949\\
250	1278229.32587288\\
300	330544.275406391\\
350	103197.272540649\\
400	32547.580252232\\
450	7754.3277374177\\
500	1669.99157737434\\
550	1455.65402022385\\
600	139.887554266909\\
650	102.003582696369\\
700	75.8091506229808\\
750	57.5104963160869\\
800	44.5391136528303\\
850	34.8167302927492\\
900	27.7417537164472\\
950	22.2854335043913\\
1000	18.1790630563774\\
};
\addlegendentry{\tiny{MmWave (30 GHz)}}

\addplot [color=NavyBlue, dash pattern={on 5pt off 3pt on 0pt off 0pt} , line width=0.7pt]
  table[row sep=crcr]{%
50	38150379.7096857\\
100	10501480.4873863\\
150	3248413.68775648\\
200	1022241.20895491\\
250	350310.052033619\\
300	113650.812096377\\
350	44299.7438894573\\
400	15566.9699388779\\
450	6531.56024748531\\
500	3797.84783996777\\
550	1646.81340637358\\
600	1069.29867061729\\
650	626.77451609647\\
700	465.754433308332\\
750	352.655632874368\\
800	273.431210528666\\
850	214.372267033299\\
900	170.411310920494\\
950	137.333373577118\\
1000	111.864244910661\\
};
\addlegendentry{\tiny{Sub-6 GHz (3 GHz)}}

\end{axis}
\end{tikzpicture}%
    \caption{Impacts of the carrier frequency.}
    \label{b1}
    \end{subfigure}
    \caption[map]{System performance. (a) Illustration of the outage floors created by the different impairment models. (b) MmWave vs Sub-6 GHz in terms of the achievable rate.}
    \label{rfidtag_testing}
\end{figure}
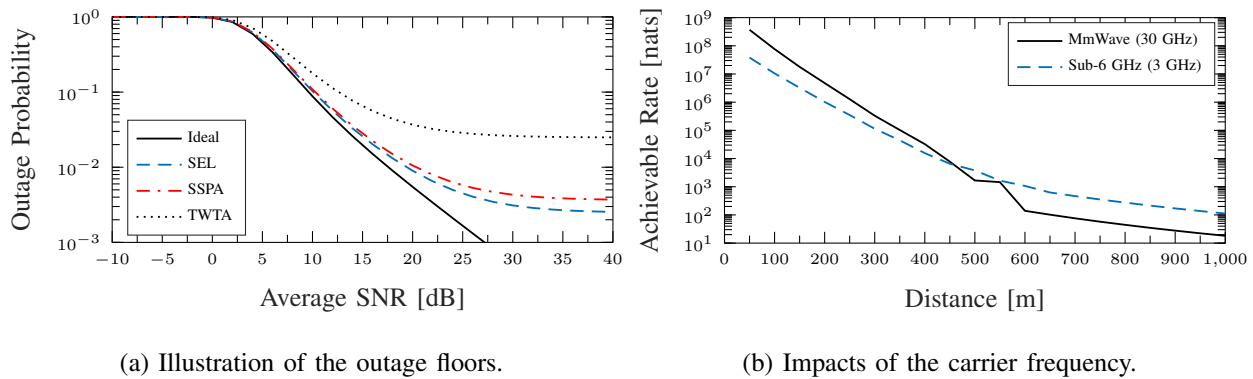
Fig.~8.a illustrates the variations of the spectral efficiency with respect to the input-back-off (IBO) level. We define the IBO as the ratio between the amplifier saturation level and the mean power of the signal $\left(\text{IBO} = \frac{A_{\text{sat}}}{\sigma_{r}}\right)$. Basically, the proposed approximation (\ref{approx}) provides aa excellent fit to the exact performance. In addition, the exact, the approximation and the Jensen's upper bound asymptotically converge to the capacity ceiling. For ideal harware, the achievable rate increases constantly with the average SNR with any constraints. This rate growth becomes linear as the average SNR becomes larger revealing that the system achieves a non-zero multiplexing gain. For the harware impairments case, the impacts of the impairments is small as the performance are perfectly aligned with the ideal hardware performance. However, the effects of the impairments become pronounced at high SNR introducing different ceilings that saturate the achievable rate. The losses are significant for lower IBO values. In fact, lower IBO value resulted from low power delivered by the amplifier to satisfy the system requirement. If the delivered power is insufficient, a detructive distortion is created and causes clipping to the signal peaks. Most importantly, the losses affected the rate are significant in the way that they completely kill the multiplexing gain of the system.
\textcolor{black}{Fig.~8.b illustrates the dependence of the coverage probability on the blockage density. We observe that for lower $\mu$ (for every $\mu$ distance, there is a blocking obstacle translated into larger blockage density), the coverage significantly degrades and conversely the probability of coverage improves for large values of $\mu$ (smaller blockage density). We also observe that the system still exhibits coverage for the SINR range between -25 dB to 5 dB for lower blockage density. This result is explained by the fact that the relative probability of LOS is still higher compared to the case of $\mu = 5 m$ where the probability of LOS is roughly null. Given that mmWaves are sensitive to blockage, reliable communications occur in LOS configuration resulting in a non-zero coverage for relatively moderate to high probability of LOS.}
\begin{figure}[htbp]
\begin{subfigure}[b]{0.5\textwidth}
\centering
\setlength\fheight{3cm}
\setlength\fwidth{7cm}
%
%
\begin{tikzpicture}

\begin{axis}[%
width=0.951\fwidth,
height=\fheight,
at={(0\fwidth,0\fheight)},
scale only axis,
xmin=-10,
xmax=30,
xlabel style={font=\color{white!15!black}},
xtick={-10,-5,0,5,10,15,20,25,30},
ytick = {0,1,2,3,4,5,6},
xlabel={\small{Average SNR [dB]}},
ymin=0,
ymax=6,
ylabel style={font=\color{white!15!black}},
ylabel={\small{Achievable Rate [nats/s/Hz]}},
axis background/.style={fill=white},
legend style={at={(0.03,0.97)}, anchor=north west, legend cell align=left, align=left, draw=white!15!black}
]
\addplot [color=black, line width=0.7pt]
  table[row sep=crcr]{%
-10	0.04695307472863\\
-8	0.0729250982348003\\
-6	0.112147088833356\\
-4	0.170126020141987\\
-2	0.253494760768546\\
0	0.369394866631294\\
2	0.524405814401161\\
4	0.723267543792237\\
6	0.967849116754414\\
8	1.25678180520644\\
10	1.58588177789294\\
12	1.94913607452535\\
14	2.33984624161859\\
16	2.75157948703051\\
18	3.17876098211665\\
20	3.61691394813636\\
22	4.06264678224161\\
24	4.51350279036302\\
26	4.96776277110577\\
28	5.4242548089115\\
30	5.88219579902107\\
};
\addlegendentry{\tiny{Exact (\ref{rate-fso})}}

\addplot [color=NavyBlue,dash pattern={on 5pt off 3pt on 0pt off 0pt} , line width=0.7pt]
  table[row sep=crcr]{%
-10	0.0474037976921395\\
-8	0.0739608241422968\\
-6	0.114425326575034\\
-4	0.174840870138708\\
-2	0.262479649072955\\
0	0.384746856146188\\
2	0.547090272003182\\
4	0.750441124958823\\
6	0.989496080078441\\
8	1.25304172936211\\
10	1.52628219471388\\
12	1.79396249020244\\
14	2.04301727530971\\
16	2.26413820029544\\
18	2.45224498064205\\
20	2.60612130036919\\
22	2.72755095981642\\
24	2.82026833366908\\
26	2.88896478340458\\
28	2.93849757341644\\
30	2.9733561334115\\
};
\addlegendentry{\tiny{Approximation (\ref{approx})}}

\addplot [color=red, dash pattern={on 5pt off 3pt on 1pt off 3pt} , line width=0.7pt]
  table[row sep=crcr]{%
-10	0.0474719421177523\\
-8	0.0741264771394335\\
-6	0.114821018806161\\
-4	0.175762119113479\\
-2	0.264548365479234\\
0	0.389172663088592\\
2	0.555998164458668\\
4	0.767124968307318\\
6	1.01834656187567\\
8	1.29888609703216\\
10	1.59303291964282\\
12	1.88282974184396\\
14	2.15103191658021\\
16	2.38399849147798\\
18	2.5740009679414\\
20	2.72001627681461\\
22	2.82651290176705\\
24	2.90094645393009\\
26	2.95131020137506\\
28	2.98460159634747\\
30	3.00625637340731\\
};
\addlegendentry{\tiny{Upper bound (\ref{jensen})}}

\addplot [color=black, dotted, line width=0.7pt]
  table[row sep=crcr]{%
-10	3.04452243772342\\
-8	3.04452243772342\\
-6	3.04452243772342\\
-4	3.04452243772342\\
-2	3.04452243772342\\
0	3.04452243772342\\
2	3.04452243772342\\
4	3.04452243772342\\
6	3.04452243772342\\
8	3.04452243772342\\
10	3.04452243772342\\
12	3.04452243772342\\
14	3.04452243772342\\
16	3.04452243772342\\
18	3.04452243772342\\
20	3.04452243772342\\
22	3.04452243772342\\
24	3.04452243772342\\
26	3.04452243772342\\
28	3.04452243772342\\
30	3.04452243772342\\
};
\addlegendentry{\tiny{Ceiling (\ref{ceiling})}}

\addplot [color=black, line width=0.7pt, forget plot]
  table[row sep=crcr]{%
-10	0.0467815084232415\\
-8	0.072517525466639\\
-6	0.111202568967792\\
-4	0.168006283737822\\
-2	0.24892019859433\\
0	0.359956139590561\\
2	0.505853155557392\\
4	0.688568923064945\\
6	0.906047211447088\\
8	1.15173822363162\\
10	1.41510063670195\\
12	1.68300323205763\\
14	1.94171715183793\\
16	2.17905390836446\\
18	2.38613490279346\\
20	2.55834028226701\\
22	2.69524437678775\\
24	2.79971006954565\\
26	2.87657414236455\\
28	2.93137780692944\\
30	2.96943006194884\\
};
\addplot [color=black, line width=0.7pt, forget plot]
  table[row sep=crcr]{%
-10	0.0462769371916542\\
-8	0.0713308474003182\\
-6	0.108491610682184\\
-4	0.162041769143071\\
-2	0.236389039785324\\
0	0.335002661758854\\
2	0.459020492077916\\
4	0.606016820282265\\
6	0.769579550115633\\
8	0.940122009035111\\
10	1.10679393594883\\
12	1.25979457905605\\
14	1.39220706868429\\
16	1.5007761806524\\
18	1.58560064384226\\
20	1.64914130495444\\
22	1.69506109796822\\
24	1.72726901057059\\
26	1.74931526845225\\
28	1.76411460395048\\
30	1.77389826750792\\
};
\addplot [color=RoyalBlue, dash pattern={on 5pt off 3pt on 0pt off 0pt} , line width=0.7pt, forget plot]
  table[row sep=crcr]{%
-10	0.0468719482134985\\
-8	0.0726794705356322\\
-6	0.111405252306635\\
-4	0.167944683229789\\
-2	0.247406863891332\\
0	0.353630732244804\\
2	0.487162370081137\\
4	0.643692785776671\\
6	0.814188016052306\\
8	0.986994614684209\\
10	1.15082219315725\\
12	1.29711020932562\\
14	1.42102989283879\\
16	1.52126883213267\\
18	1.59914239255403\\
20	1.65753616956824\\
22	1.6999871288239\\
24	1.73002813472779\\
26	1.75080162501224\\
28	1.76488981111655\\
30	1.77429205576855\\
};
\addplot [color=red, dash pattern={on 5pt off 3pt on 1pt off 3pt} , line width=0.7pt, forget plot]
  table[row sep=crcr]{%
-10	0.0471362322513366\\
-8	0.0733109653895825\\
-6	0.112874857329987\\
-4	0.171236257866285\\
-2	0.254396591575648\\
0	0.367464721427276\\
2	0.512252807887739\\
4	0.684813956757922\\
6	0.874547730352661\\
8	1.06609579820706\\
10	1.24357117708317\\
12	1.39505872297225\\
14	1.51513390674619\\
16	1.60452336707743\\
18	1.66783824455253\\
20	1.71104748517056\\
22	1.73976632733377\\
24	1.75851192160295\\
26	1.77060112576392\\
28	1.77833638903546\\
30	1.78326068458893\\
};
\addplot [color=black, dotted, line width=0.7pt, forget plot]
  table[row sep=crcr]{%
-10	1.79175946922806\\
-8	1.79175946922806\\
-6	1.79175946922806\\
-4	1.79175946922806\\
-2	1.79175946922806\\
0	1.79175946922806\\
2	1.79175946922806\\
4	1.79175946922806\\
6	1.79175946922806\\
8	1.79175946922806\\
10	1.79175946922806\\
12	1.79175946922806\\
14	1.79175946922806\\
16	1.79175946922806\\
18	1.79175946922806\\
20	1.79175946922806\\
22	1.79175946922806\\
24	1.79175946922806\\
26	1.79175946922806\\
28	1.79175946922806\\
30	1.79175946922806\\
};
\node[right, align=left]
at (axis cs:16.8,5) {\tiny{Ideal hardware}};
\node[right, align=left]
at (axis cs:21.6,3.3) {\tiny{IBO = 2 dB}};
\node[right, align=left]
at (axis cs:21.6,2.1) {\tiny{IBO = 1 dB}};
\end{axis}
\end{tikzpicture}%
    \caption{Effects of the IBO factor.}
    \label{a1}
    \end{subfigure}
    \begin{subfigure}[b]{0.5\textwidth}
\centering
\setlength\fheight{3cm}
\setlength\fwidth{7cm}
%
%
\begin{tikzpicture}

\begin{axis}[%
width=0.951\fwidth,
height=\fheight,
at={(0\fwidth,0\fheight)},
scale only axis,
xmin=-40,
xmax=20,
xlabel style={font=\color{white!15!black}},
xlabel={\small{SINR Threshold [dB]}},
xtick={-40,-35,-30,-25,-20,-15,-10,-5,0,5,10,15,20},
ymin=0,
ymax=1,
ylabel style={font=\color{white!15!black}},
ylabel={\small{Coverage Probability}},
axis background/.style={fill=white},
legend style={legend cell align=left, align=left, draw=white!15!black}
]
\addplot [color=black, line width=0.7pt]
  table[row sep=crcr]{%
-40	1\\
-39	1\\
-38	1\\
-37	0.99999\\
-36	0.99977\\
-35	0.99814\\
-34	0.99154\\
-33	0.97162\\
-32	0.92785\\
-31	0.8543\\
-30	0.75254\\
-29	0.6318\\
-28	0.50966\\
-27	0.39509\\
-26	0.2944\\
-25	0.21357\\
-24	0.15112\\
-23	0.10438\\
-22	0.06964\\
-21	0.0463\\
-20	0.03043\\
-19	0.0196\\
-18	0.01278\\
-17	0.00812999999999997\\
-16	0.00517999999999996\\
-15	0.00341000000000002\\
-14	0.00204000000000004\\
-13	0.00131000000000003\\
-12	0.000839999999999952\\
-11	0.000539999999999985\\
-10	0.000290000000000012\\
-9	0.000179999999999958\\
-8	0.000110000000000054\\
-7	4.99999999999945e-05\\
-6	9.99999999995449e-06\\
-5	0\\
-4	0\\
-3	0\\
-2	9.99999999995449e-06\\
-1	9.99999999995449e-06\\
0	0\\
1	9.99999999995449e-06\\
2	9.99999999995449e-06\\
3	9.99999999995449e-06\\
4	0\\
5	0\\
6	0\\
7	0\\
8	9.99999999995449e-06\\
9	0\\
10	0\\
11	0\\
12	9.99999999995449e-06\\
13	0\\
14	0\\
15	0\\
16	0\\
17	0\\
18	0\\
19	0\\
20	0\\
};
\addlegendentry{\tiny{$\mu\text{ = 5 m}$}}

\addplot [color=NavyBlue, dash pattern={on 5pt off 3pt on 0pt off 0pt} , line width=0.7pt]
  table[row sep=crcr]{%
-40	1\\
-39	1\\
-38	1\\
-37	0.99999\\
-36	0.99978\\
-35	0.99827\\
-34	0.99198\\
-33	0.97318\\
-32	0.9316\\
-31	0.86193\\
-30	0.76503\\
-29	0.6503\\
-28	0.53337\\
-27	0.42475\\
-26	0.32964\\
-25	0.25246\\
-24	0.19246\\
-23	0.14878\\
-22	0.11466\\
-21	0.09434\\
-20	0.07772\\
-19	0.06748\\
-18	0.06209\\
-17	0.05715\\
-16	0.0545099999999999\\
-15	0.05169\\
-14	0.05132\\
-13	0.0513400000000001\\
-12	0.05017\\
-11	0.05033\\
-10	0.04972\\
-9	0.0512\\
-8	0.05045\\
-7	0.05015\\
-6	0.04902\\
-5	0.0488\\
-4	0.04805\\
-3	0.04889\\
-2	0.05053\\
-1	0.04942\\
0	0.05039\\
1	0.04999\\
2	0.04842\\
3	0.04866\\
4	0.04435\\
5	0.03938\\
6	0.03526\\
7	0.0284799999999999\\
8	0.02185\\
9	0.01617\\
10	0.0124300000000001\\
11	0.00910999999999995\\
12	0.00666\\
13	0.00448000000000004\\
14	0.00304000000000004\\
15	0.00197999999999998\\
16	0.00109999999999999\\
17	0.000780000000000003\\
18	0.00048999999999999\\
19	0.000249999999999972\\
20	0.000219999999999998\\
};
\addlegendentry{\tiny{$\mu\text{ = 20 m}$}}

\addplot [color=red,dash pattern={on 5pt off 3pt on 1pt off 3pt} , line width=0.7pt]
  table[row sep=crcr]{%
-40	1\\
-39	1\\
-38	1\\
-37	0.99999\\
-36	0.99981\\
-35	0.9984\\
-34	0.99259\\
-33	0.97537\\
-32	0.93776\\
-31	0.87445\\
-30	0.78535\\
-29	0.68095\\
-28	0.5754\\
-27	0.47686\\
-26	0.38897\\
-25	0.31976\\
-24	0.26563\\
-23	0.22471\\
-22	0.19554\\
-21	0.17566\\
-20	0.16388\\
-19	0.15364\\
-18	0.14604\\
-17	0.14253\\
-16	0.13839\\
-15	0.13945\\
-14	0.13792\\
-13	0.13628\\
-12	0.13637\\
-11	0.13664\\
-10	0.13438\\
-9	0.13441\\
-8	0.13561\\
-7	0.13515\\
-6	0.13458\\
-5	0.13444\\
-4	0.13473\\
-3	0.13639\\
-2	0.13599\\
-1	0.13635\\
0	0.13447\\
1	0.13528\\
2	0.13316\\
3	0.13009\\
4	0.12206\\
5	0.11045\\
6	0.09468\\
7	0.07865\\
8	0.0617\\
9	0.04641\\
10	0.0348000000000001\\
11	0.02472\\
12	0.01745\\
13	0.0112100000000001\\
14	0.00736999999999999\\
15	0.00504000000000004\\
16	0.00327\\
17	0.00210999999999995\\
18	0.00151000000000001\\
19	0.000829999999999997\\
20	0.000770000000000048\\
};
\addlegendentry{\tiny{$\mu\text{ = 30 m}$}}

\addplot [color=black, dotted, line width=0.7pt]
  table[row sep=crcr]{%
-40	1\\
-39	1\\
-38	1\\
-37	0.99999\\
-36	0.99981\\
-35	0.99864\\
-34	0.99345\\
-33	0.97792\\
-32	0.9439\\
-31	0.88641\\
-30	0.80783\\
-29	0.71366\\
-28	0.61869\\
-27	0.53053\\
-26	0.45079\\
-25	0.38953\\
-24	0.34141\\
-23	0.30498\\
-22	0.27922\\
-21	0.26049\\
-20	0.24897\\
-19	0.23808\\
-18	0.23232\\
-17	0.23041\\
-16	0.22837\\
-15	0.22478\\
-14	0.22318\\
-13	0.22225\\
-12	0.22144\\
-11	0.22373\\
-10	0.22371\\
-9	0.22283\\
-8	0.22302\\
-7	0.22326\\
-6	0.2229\\
-5	0.22281\\
-4	0.22121\\
-3	0.22477\\
-2	0.22465\\
-1	0.22163\\
0	0.22047\\
1	0.22199\\
2	0.22048\\
3	0.21271\\
4	0.19831\\
5	0.18132\\
6	0.15615\\
7	0.12868\\
8	0.1024\\
9	0.07713\\
10	0.05677\\
11	0.04176\\
12	0.0274799999999999\\
13	0.01956\\
14	0.01266\\
15	0.00897000000000003\\
16	0.00541999999999998\\
17	0.00348999999999999\\
18	0.00219000000000003\\
19	0.00141999999999998\\
20	0.000839999999999952\\
};
\addlegendentry{\tiny{$\mu\text{ = 40 m}$}}

\end{axis}
\end{tikzpicture}%
    \caption{Impacts of the blockage density.}
    \label{b1}
    \end{subfigure}
    \caption[map]{System performance. (a) Exact, approximation, upper bound, and ceiling of the achievable rate with respect to the average SNR. (b) Probability of coverage dependence on different blockage densities. }
    \label{rfidtag_testing}
\end{figure}
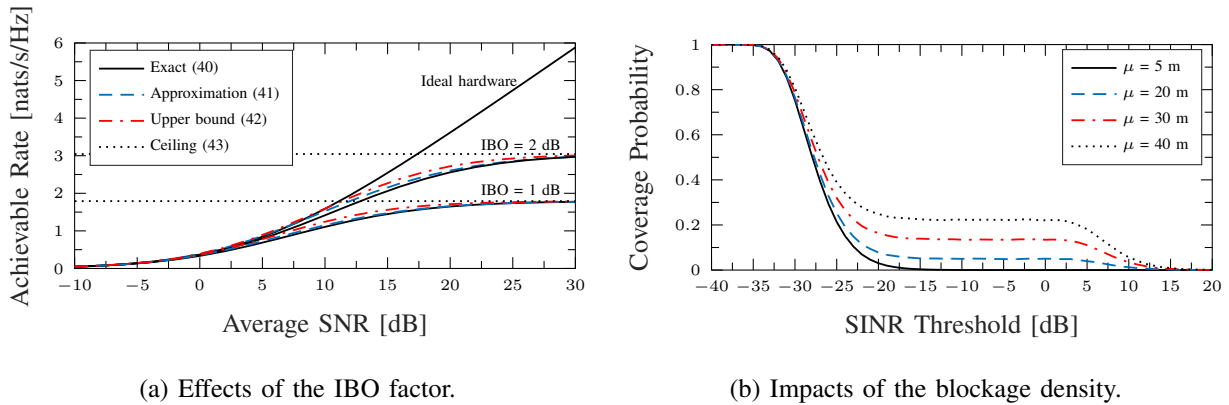
\textcolor{black}{Fig.~9.a presents the variations of the rate coverage with respect to the average number of interferers. As expected, the system achieves an acceptable rate coverage with minimum number of interferers and conversely the coverage deteriorates as the interference density becomes larger}.
\textcolor{black}{Fig.~9.b illustrates the effects of the NLOS pathloss exponent on the rate coverage. For this scenario we consider a low probability of LOS ($p_{\text{los}} = 0.1$) to allow for the NLOS scenario to occur with high probability. We observe for a lower NLOS pathloss exponent, the system can achieve higher target rates in the order of $2\times10^9$ nats with a probability higher than 0.75. However, as the NLOS pathloss exponent becomes more severe ($\alpha_{\text{nlos}} = 3.5$), the rate coverage decreases and the system certainly cannot achieve a target rate higher than $1.5\times10^9$ nats}.

\begin{figure}[htbp]
\begin{subfigure}[b]{0.5\textwidth}
\centering
\setlength\fheight{3cm}
\setlength\fwidth{7cm}
%
%
\begin{tikzpicture}

\begin{axis}[%
width=0.951\fwidth,
height=\fheight,
at={(0\fwidth,0\fheight)},
scale only axis,
xmin=0,
xmax=3000000000,
xtick = {0,0.25e9,0.5e9,0.75e9,1e9,1.25e9,1.5e9,1.75e9,2e9,2.25e9,2.5e9,2.75e9,3e9},
xlabel style={font=\color{white!15!black}},
xlabel={\small{Target Rate [nats]}},
ymin=0,
ymax=1,
ylabel style={font=\color{white!15!black}},
ylabel={\small{Rate Coverage}},
axis background/.style={fill=white},
legend style={legend cell align=left, align=left, draw=white!15!black}
]
\addplot [color=black, line width=0.7pt]
  table[row sep=crcr]{%
0	1\\
30000000	1\\
60000000	1\\
90000000	1\\
120000000	1\\
150000000	1\\
180000000	1\\
210000000	1\\
240000000	1\\
270000000	1\\
300000000	1\\
330000000	1\\
360000000	1\\
390000000	1\\
420000000	1\\
450000000	1\\
480000000	1\\
510000000	1\\
540000000	0.9996\\
570000000	0.9988\\
600000000	0.9977\\
630000000	0.9946\\
660000000	0.9901\\
690000000	0.9841\\
720000000	0.9748\\
750000000	0.9629\\
780000000	0.9485\\
810000000	0.93\\
840000000	0.9082\\
870000000	0.882\\
900000000	0.8551\\
930000000	0.8255\\
960000000	0.7883\\
990000000	0.7538\\
1020000000	0.7169\\
1050000000	0.6783\\
1080000000	0.6408\\
1110000000	0.603\\
1140000000	0.5678\\
1170000000	0.5323\\
1200000000	0.4979\\
1230000000	0.463\\
1260000000	0.4265\\
1290000000	0.3943\\
1320000000	0.3616\\
1350000000	0.3308\\
1380000000	0.3057\\
1410000000	0.2831\\
1440000000	0.2591\\
1470000000	0.2354\\
1500000000	0.2185\\
1530000000	0.1986\\
1560000000	0.1812\\
1590000000	0.165\\
1620000000	0.1504\\
1650000000	0.1362\\
1680000000	0.1248\\
1710000000	0.1125\\
1740000000	0.1034\\
1770000000	0.0938\\
1800000000	0.0869\\
1830000000	0.081\\
1860000000	0.0736\\
1890000000	0.0678\\
1920000000	0.0627\\
1950000000	0.0568\\
1980000000	0.0514\\
2010000000	0.0477\\
2040000000	0.0431\\
2070000000	0.0392\\
2100000000	0.036\\
2130000000	0.0346\\
2160000000	0.0319\\
2190000000	0.0295\\
2220000000	0.0256\\
2250000000	0.0239\\
2280000000	0.0213\\
2310000000	0.0192\\
2340000000	0.0165\\
2370000000	0.0163\\
2400000000	0.0142\\
2430000000	0.0142\\
2460000000	0.0123\\
2490000000	0.0115\\
2520000000	0.0094\\
2550000000	0.0086\\
2580000000	0.0081\\
2610000000	0.0079\\
2640000000	0.0072\\
2670000000	0.0074\\
2700000000	0.007\\
2730000000	0.0057\\
2760000000	0.0059\\
2790000000	0.006\\
2820000000	0.0057\\
2850000000	0.0063\\
2880000000	0.0044\\
2910000000	0.0037\\
2940000000	0.0033\\
2970000000	0.0044\\
3000000000	0.0039\\
};
\addlegendentry{\tiny{$M_z = 1$}}

\addplot [color=NavyBlue, dash pattern={on 5pt off 3pt on 0pt off 0pt} , line width=0.7pt]
  table[row sep=crcr]{%
0	1\\
30000000	1\\
60000000	1\\
90000000	1\\
120000000	1\\
150000000	1\\
180000000	1\\
210000000	1\\
240000000	1\\
270000000	1\\
300000000	1\\
330000000	1\\
360000000	0.9993\\
390000000	0.9974\\
420000000	0.9923\\
450000000	0.9786\\
480000000	0.952\\
510000000	0.9082\\
540000000	0.8477\\
570000000	0.7677\\
600000000	0.6733\\
630000000	0.576\\
660000000	0.4758\\
690000000	0.3837\\
720000000	0.3029\\
750000000	0.2349\\
780000000	0.174\\
810000000	0.1281\\
840000000	0.0937\\
870000000	0.0662\\
900000000	0.0462\\
930000000	0.0326\\
960000000	0.0224\\
990000000	0.0153\\
1020000000	0.011\\
1050000000	0.0076\\
1080000000	0.0054\\
1110000000	0.0041\\
1140000000	0.0033\\
1170000000	0.0037\\
1200000000	0.0038\\
1230000000	0.0022\\
1260000000	0.0032\\
1290000000	0.0027\\
1320000000	0.0026\\
1350000000	0.0038\\
1380000000	0.0025\\
1410000000	0.0027\\
1440000000	0.0033\\
1470000000	0.0021\\
1500000000	0.0014\\
1530000000	0.0021\\
1560000000	0.0023\\
1590000000	0.0036\\
1620000000	0.0033\\
1650000000	0.0025\\
1680000000	0.0027\\
1710000000	0.0028\\
1740000000	0.0031\\
1770000000	0.0017\\
1800000000	0.0021\\
1830000000	0.0018\\
1860000000	0.0019\\
1890000000	0.0026\\
1920000000	0.0026\\
1950000000	0.0028\\
1980000000	0.0018\\
2010000000	0.0024\\
2040000000	0.0027\\
2070000000	0.0025\\
2100000000	0.002\\
2130000000	0.0018\\
2160000000	0.0022\\
2190000000	0.0026\\
2220000000	0.002\\
2250000000	0.0025\\
2280000000	0.0025\\
2310000000	0.0019\\
2340000000	0.0024\\
2370000000	0.0031\\
2400000000	0.0028\\
2430000000	0.0019\\
2460000000	0.0021\\
2490000000	0.0026\\
2520000000	0.0025\\
2550000000	0.0024\\
2580000000	0.0027\\
2610000000	0.0024\\
2640000000	0.0021\\
2670000000	0.0026\\
2700000000	0.0027\\
2730000000	0.0024\\
2760000000	0.0032\\
2790000000	0.0021\\
2820000000	0.0023\\
2850000000	0.0021\\
2880000000	0.0028\\
2910000000	0.0017\\
2940000000	0.0012\\
2970000000	0.0026\\
3000000000	0.0017\\
};
\addlegendentry{\tiny{$M_z = 5$}}

\addplot [color=red, dash pattern={on 5pt off 3pt on 1pt off 3pt} , line width=0.7pt]
  table[row sep=crcr]{%
0	1\\
30000000	1\\
60000000	1\\
90000000	1\\
120000000	1\\
150000000	1\\
180000000	1\\
210000000	1\\
240000000	1\\
270000000	0.9999\\
300000000	0.9979\\
330000000	0.9857\\
360000000	0.9465\\
390000000	0.866\\
420000000	0.7411\\
450000000	0.5895\\
480000000	0.4305\\
510000000	0.3019\\
540000000	0.1992\\
570000000	0.1179\\
600000000	0.0686\\
630000000	0.0367\\
660000000	0.0194\\
690000000	0.0117\\
720000000	0.0068\\
750000000	0.0052\\
780000000	0.0034\\
810000000	0.0028\\
840000000	0.0025\\
870000000	0.0021\\
900000000	0.0027\\
930000000	0.0027\\
960000000	0.0035\\
990000000	0.002\\
1020000000	0.0016\\
1050000000	0.0019\\
1080000000	0.0016\\
1110000000	0.0022\\
1140000000	0.0019\\
1170000000	0.0029\\
1200000000	0.0029\\
1230000000	0.0031\\
1260000000	0.0019\\
1290000000	0.0024\\
1320000000	0.0026\\
1350000000	0.0028\\
1380000000	0.003\\
1410000000	0.0022\\
1440000000	0.0021\\
1470000000	0.0028\\
1500000000	0.0032\\
1530000000	0.0022\\
1560000000	0.0027\\
1590000000	0.0027\\
1620000000	0.0017\\
1650000000	0.002\\
1680000000	0.0027\\
1710000000	0.0024\\
1740000000	0.0026\\
1770000000	0.003\\
1800000000	0.0022\\
1830000000	0.0038\\
1860000000	0.0022\\
1890000000	0.004\\
1920000000	0.0026\\
1950000000	0.0028\\
1980000000	0.0023\\
2010000000	0.0026\\
2040000000	0.0024\\
2070000000	0.002\\
2100000000	0.0022\\
2130000000	0.0024\\
2160000000	0.0023\\
2190000000	0.0026\\
2220000000	0.0017\\
2250000000	0.0026\\
2280000000	0.0027\\
2310000000	0.0024\\
2340000000	0.0019\\
2370000000	0.0024\\
2400000000	0.0023\\
2430000000	0.0032\\
2460000000	0.0023\\
2490000000	0.0027\\
2520000000	0.003\\
2550000000	0.0018\\
2580000000	0.0016\\
2610000000	0.0024\\
2640000000	0.0021\\
2670000000	0.0025\\
2700000000	0.0022\\
2730000000	0.0017\\
2760000000	0.0019\\
2790000000	0.0013\\
2820000000	0.0022\\
2850000000	0.0017\\
2880000000	0.002\\
2910000000	0.001\\
2940000000	0.0004\\
2970000000	0.0003\\
3000000000	0.0003\\
};
\addlegendentry{\tiny{$M_z = 10$}}

\addplot [color=black, dotted, line width=0.7pt]
  table[row sep=crcr]{%
0	1\\
30000000	1\\
60000000	1\\
90000000	1\\
120000000	1\\
150000000	1\\
180000000	1\\
210000000	0.9983\\
240000000	0.9734\\
270000000	0.8558\\
300000000	0.6181\\
330000000	0.3544\\
360000000	0.1621\\
390000000	0.062\\
420000000	0.0226\\
450000000	0.0069\\
480000000	0.0047\\
510000000	0.0031\\
540000000	0.0029\\
570000000	0.002\\
600000000	0.0025\\
630000000	0.0025\\
660000000	0.0024\\
690000000	0.0024\\
720000000	0.0037\\
750000000	0.0021\\
780000000	0.0029\\
810000000	0.0034\\
840000000	0.0029\\
870000000	0.0021\\
900000000	0.003\\
930000000	0.0034\\
960000000	0.0024\\
990000000	0.0021\\
1020000000	0.0025\\
1050000000	0.0025\\
1080000000	0.0025\\
1110000000	0.0021\\
1140000000	0.0027\\
1170000000	0.0018\\
1200000000	0.0016\\
1230000000	0.0031\\
1260000000	0.0024\\
1290000000	0.0019\\
1320000000	0.0023\\
1350000000	0.0031\\
1380000000	0.0032\\
1410000000	0.003\\
1440000000	0.0025\\
1470000000	0.0036\\
1500000000	0.0023\\
1530000000	0.0022\\
1560000000	0.0027\\
1590000000	0.0021\\
1620000000	0.0027\\
1650000000	0.0029\\
1680000000	0.0024\\
1710000000	0.0018\\
1740000000	0.0026\\
1770000000	0.0026\\
1800000000	0.0024\\
1830000000	0.0027\\
1860000000	0.0028\\
1890000000	0.0017\\
1920000000	0.0025\\
1950000000	0.002\\
1980000000	0.0036\\
2010000000	0.0015\\
2040000000	0.0029\\
2070000000	0.003\\
2100000000	0.0026\\
2130000000	0.0026\\
2160000000	0.0015\\
2190000000	0.0032\\
2220000000	0.0023\\
2250000000	0.0025\\
2280000000	0.0016\\
2310000000	0.0024\\
2340000000	0.0028\\
2370000000	0.0029\\
2400000000	0.0017\\
2430000000	0.002\\
2460000000	0.0022\\
2490000000	0.0024\\
2520000000	0.0028\\
2550000000	0.0021\\
2580000000	0.0013\\
2610000000	0.0016\\
2640000000	0.0018\\
2670000000	0.0005\\
2700000000	0.0004\\
2730000000	0.0002\\
2760000000	0.0001\\
2790000000	0.0001\\
2820000000	0\\
2850000000	0\\
2880000000	0\\
2910000000	0\\
2940000000	0\\
2970000000	0\\
3000000000	0\\
};
\addlegendentry{\tiny{$M_z = 20$}}

\end{axis}
\end{tikzpicture}%
    \caption{Effects of the average number of the interferers.}
    \label{a1}
    \end{subfigure}
    \begin{subfigure}[b]{0.5\textwidth}
\centering
\setlength\fheight{3cm}
\setlength\fwidth{7cm}
%
%
\begin{tikzpicture}

\begin{axis}[%
width=0.951\fwidth,
height=\fheight,
at={(0\fwidth,0\fheight)},
scale only axis,
xmin=0,
xmax=3000000000,
xtick = {0,0.25e9,0.5e9,0.75e9,1e9,1.25e9,1.5e9,1.75e9,2e9,2.25e9,2.5e9,2.75e9,3e9},
xlabel style={font=\color{white!15!black}},
xlabel={\small{Target Rate [nats]}},
ymin=0,
ymax=1,
ylabel style={font=\color{white!15!black}},
ylabel={\small{Rate Coverage}},
axis background/.style={fill=white},
legend style={at={(0.03,0.03)}, anchor=south west, legend cell align=left, align=left, draw=white!15!black}
]
\addplot [color=black, line width=0.7pt]
  table[row sep=crcr]{%
0	1\\
30000000	1\\
60000000	1\\
90000000	1\\
120000000	1\\
150000000	1\\
180000000	1\\
210000000	1\\
240000000	1\\
270000000	1\\
300000000	1\\
330000000	1\\
360000000	1\\
390000000	1\\
420000000	1\\
450000000	1\\
480000000	1\\
510000000	1\\
540000000	1\\
570000000	1\\
600000000	1\\
630000000	1\\
660000000	1\\
690000000	1\\
720000000	1\\
750000000	1\\
780000000	1\\
810000000	1\\
840000000	1\\
870000000	1\\
900000000	1\\
930000000	1\\
960000000	1\\
990000000	1\\
1020000000	1\\
1050000000	1\\
1080000000	1\\
1110000000	1\\
1140000000	1\\
1170000000	1\\
1200000000	1\\
1230000000	1\\
1260000000	1\\
1290000000	1\\
1320000000	1\\
1350000000	1\\
1380000000	1\\
1410000000	1\\
1440000000	1\\
1470000000	1\\
1500000000	1\\
1530000000	0.9998\\
1560000000	0.9987\\
1590000000	0.9977\\
1620000000	0.9952\\
1650000000	0.9909\\
1680000000	0.9865\\
1710000000	0.9775\\
1740000000	0.9635\\
1770000000	0.9424\\
1800000000	0.9188\\
1830000000	0.8888\\
1860000000	0.8568\\
1890000000	0.8171\\
1920000000	0.7736\\
1950000000	0.7221\\
1980000000	0.6703\\
2010000000	0.6165\\
2040000000	0.5617\\
2070000000	0.5058\\
2100000000	0.4546\\
2130000000	0.4048\\
2160000000	0.359\\
2190000000	0.3153\\
2220000000	0.2738\\
2250000000	0.243\\
2280000000	0.2129\\
2310000000	0.1827\\
2340000000	0.1577\\
2370000000	0.1339\\
2400000000	0.1131\\
2430000000	0.0967\\
2460000000	0.0837\\
2490000000	0.0698\\
2520000000	0.0596\\
2550000000	0.0497\\
2580000000	0.0416\\
2610000000	0.0354\\
2640000000	0.03\\
2670000000	0.0264\\
2700000000	0.022\\
2730000000	0.0186\\
2760000000	0.0152\\
2790000000	0.0119\\
2820000000	0.0097\\
2850000000	0.0078\\
2880000000	0.0067\\
2910000000	0.0056\\
2940000000	0.0048\\
2970000000	0.0039\\
3000000000	0.0036\\
};
\addlegendentry{\tiny{$\alpha{}_{\text{nlos}}\text{ = 2.5}$}}

\addplot [color=NavyBlue, dash pattern={on 5pt off 3pt on 0pt off 0pt} , line width=0.7pt]
  table[row sep=crcr]{%
0	1\\
30000000	1\\
60000000	1\\
90000000	1\\
120000000	1\\
150000000	1\\
180000000	1\\
210000000	1\\
240000000	1\\
270000000	1\\
300000000	1\\
330000000	1\\
360000000	1\\
390000000	1\\
420000000	1\\
450000000	1\\
480000000	1\\
510000000	1\\
540000000	1\\
570000000	1\\
600000000	1\\
630000000	1\\
660000000	1\\
690000000	1\\
720000000	1\\
750000000	1\\
780000000	1\\
810000000	1\\
840000000	1\\
870000000	1\\
900000000	1\\
930000000	1\\
960000000	1\\
990000000	1\\
1020000000	1\\
1050000000	1\\
1080000000	1\\
1110000000	1\\
1140000000	1\\
1170000000	0.9997\\
1200000000	0.9986\\
1230000000	0.9975\\
1260000000	0.9942\\
1290000000	0.99\\
1320000000	0.9853\\
1350000000	0.975\\
1380000000	0.96\\
1410000000	0.9384\\
1440000000	0.9124\\
1470000000	0.8817\\
1500000000	0.8473\\
1530000000	0.8069\\
1560000000	0.7605\\
1590000000	0.7092\\
1620000000	0.662\\
1650000000	0.6019\\
1680000000	0.5458\\
1710000000	0.4911\\
1740000000	0.4423\\
1770000000	0.3923\\
1800000000	0.3463\\
1830000000	0.3034\\
1860000000	0.2638\\
1890000000	0.2346\\
1920000000	0.2049\\
1950000000	0.1759\\
1980000000	0.1497\\
2010000000	0.1277\\
2040000000	0.1083\\
2070000000	0.0929\\
2100000000	0.0792\\
2130000000	0.0672\\
2160000000	0.057\\
2190000000	0.048\\
2220000000	0.0402\\
2250000000	0.0338\\
2280000000	0.0305\\
2310000000	0.026\\
2340000000	0.0231\\
2370000000	0.019\\
2400000000	0.0149\\
2430000000	0.0124\\
2460000000	0.0115\\
2490000000	0.0087\\
2520000000	0.0081\\
2550000000	0.0072\\
2580000000	0.0057\\
2610000000	0.005\\
2640000000	0.0047\\
2670000000	0.0033\\
2700000000	0.0026\\
2730000000	0.0023\\
2760000000	0.0028\\
2790000000	0.0018\\
2820000000	0.0024\\
2850000000	0.0016\\
2880000000	0.0017\\
2910000000	0.0012\\
2940000000	0.0012\\
2970000000	0.0009\\
3000000000	0.0008\\
};
\addlegendentry{\tiny{$\alpha{}_{\text{nlos}}\text{ = 2.8}$}}

\addplot [color=red, dash pattern={on 5pt off 3pt on 1pt off 3pt} , line width=0.7pt]
  table[row sep=crcr]{%
0	1\\
30000000	1\\
60000000	1\\
90000000	1\\
120000000	1\\
150000000	1\\
180000000	1\\
210000000	1\\
240000000	1\\
270000000	1\\
300000000	1\\
330000000	1\\
360000000	1\\
390000000	1\\
420000000	1\\
450000000	1\\
480000000	1\\
510000000	1\\
540000000	1\\
570000000	1\\
600000000	1\\
630000000	1\\
660000000	1\\
690000000	1\\
720000000	0.9992\\
750000000	0.9979\\
780000000	0.9953\\
810000000	0.9908\\
840000000	0.9858\\
870000000	0.9761\\
900000000	0.9606\\
930000000	0.9379\\
960000000	0.9111\\
990000000	0.8797\\
1020000000	0.8431\\
1050000000	0.8015\\
1080000000	0.752\\
1110000000	0.7013\\
1140000000	0.6499\\
1170000000	0.5899\\
1200000000	0.5347\\
1230000000	0.4788\\
1260000000	0.4305\\
1290000000	0.378\\
1320000000	0.3342\\
1350000000	0.2922\\
1380000000	0.2546\\
1410000000	0.2267\\
1440000000	0.1948\\
1470000000	0.1662\\
1500000000	0.144\\
1530000000	0.1194\\
1560000000	0.1016\\
1590000000	0.0886\\
1620000000	0.0746\\
1650000000	0.0632\\
1680000000	0.0544\\
1710000000	0.0448\\
1740000000	0.0389\\
1770000000	0.0325\\
1800000000	0.0293\\
1830000000	0.0252\\
1860000000	0.0215\\
1890000000	0.0176\\
1920000000	0.0138\\
1950000000	0.0114\\
1980000000	0.0103\\
2010000000	0.0096\\
2040000000	0.0074\\
2070000000	0.0081\\
2100000000	0.0058\\
2130000000	0.0053\\
2160000000	0.0051\\
2190000000	0.0044\\
2220000000	0.0046\\
2250000000	0.0041\\
2280000000	0.0033\\
2310000000	0.0033\\
2340000000	0.0036\\
2370000000	0.0029\\
2400000000	0.0034\\
2430000000	0.0038\\
2460000000	0.0027\\
2490000000	0.003\\
2520000000	0.0016\\
2550000000	0.0022\\
2580000000	0.0024\\
2610000000	0.0012\\
2640000000	0.0009\\
2670000000	0.0011\\
2700000000	0.0015\\
2730000000	0.0016\\
2760000000	0.0007\\
2790000000	0.0015\\
2820000000	0.0007\\
2850000000	0.0006\\
2880000000	0.0008\\
2910000000	0.0004\\
2940000000	0.0008\\
2970000000	0.0001\\
3000000000	0\\
};
\addlegendentry{\tiny{$\alpha{}_{\text{nlos}}\text{ = 3.2}$}}

\addplot [color=black, dotted, line width=0.7pt]
  table[row sep=crcr]{%
0	1\\
30000000	1\\
60000000	1\\
90000000	1\\
120000000	1\\
150000000	1\\
180000000	1\\
210000000	1\\
240000000	1\\
270000000	1\\
300000000	1\\
330000000	1\\
360000000	1\\
390000000	0.9999\\
420000000	0.9986\\
450000000	0.9964\\
480000000	0.992\\
510000000	0.9868\\
540000000	0.9751\\
570000000	0.9565\\
600000000	0.9319\\
630000000	0.8996\\
660000000	0.8629\\
690000000	0.8206\\
720000000	0.7732\\
750000000	0.7164\\
780000000	0.6638\\
810000000	0.6018\\
840000000	0.5415\\
870000000	0.4845\\
900000000	0.4345\\
930000000	0.3807\\
960000000	0.3339\\
990000000	0.291\\
1020000000	0.2536\\
1050000000	0.2236\\
1080000000	0.1926\\
1110000000	0.1646\\
1140000000	0.1403\\
1170000000	0.1179\\
1200000000	0.0991\\
1230000000	0.0863\\
1260000000	0.0731\\
1290000000	0.063\\
1320000000	0.053\\
1350000000	0.044\\
1380000000	0.0379\\
1410000000	0.0319\\
1440000000	0.0276\\
1470000000	0.0241\\
1500000000	0.02\\
1530000000	0.0176\\
1560000000	0.0137\\
1590000000	0.0122\\
1620000000	0.01\\
1650000000	0.0094\\
1680000000	0.0078\\
1710000000	0.0072\\
1740000000	0.0066\\
1770000000	0.0055\\
1800000000	0.0045\\
1830000000	0.0046\\
1860000000	0.004\\
1890000000	0.0041\\
1920000000	0.0039\\
1950000000	0.0046\\
1980000000	0.0041\\
2010000000	0.0036\\
2040000000	0.0026\\
2070000000	0.0032\\
2100000000	0.0029\\
2130000000	0.0032\\
2160000000	0.0018\\
2190000000	0.0033\\
2220000000	0.0023\\
2250000000	0.0034\\
2280000000	0.0023\\
2310000000	0.0024\\
2340000000	0.0019\\
2370000000	0.002\\
2400000000	0.002\\
2430000000	0.0019\\
2460000000	0.0015\\
2490000000	0.0013\\
2520000000	0.0023\\
2550000000	0.0017\\
2580000000	0.0016\\
2610000000	0.002\\
2640000000	0.0013\\
2670000000	0.0011\\
2700000000	0.0014\\
2730000000	0.0014\\
2760000000	0.0009\\
2790000000	0.0005\\
2820000000	0.0009\\
2850000000	0.0006\\
2880000000	0.0012\\
2910000000	0.0007\\
2940000000	0.0004\\
2970000000	0.0002\\
3000000000	0.0004\\
};
\addlegendentry{\tiny{$\alpha{}_{\text{nlos}}\text{ = 3.5}$}}

\end{axis}
\end{tikzpicture}%
    \caption{Effects of the NLOS pathloss exponent.}
    \label{b1}
    \end{subfigure}
    \caption[map]{Rate coverage performance. (a) Illustration of the rate coverage dependence on the number of the interferers. (b) Probability of rate coverage with respect to a range of target rates for different NLOS pathloss exponent values. }
    \label{rfidtag_testing}
\end{figure}
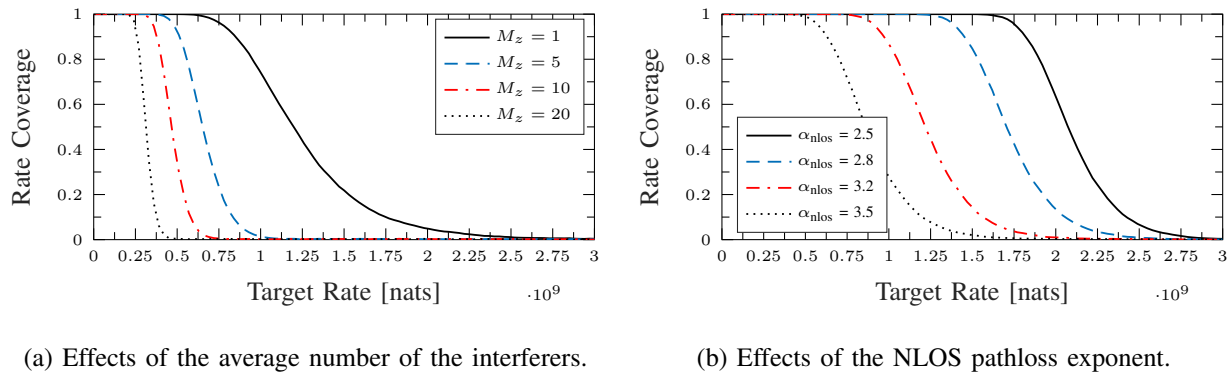

\section{Conclusion}
\textcolor{black}{In this paper, we present a tractable performance analysis of a mmWaves cellular network with FSO backhauling. We derived the closed-forms, approximations, and upper bounds of the achievable rate, the error performance, and the probability of outage. We demonstrate that the system performance depends to a large extent to the correlation between the CSIs. For full correlation the system achieves better performance, however, the performance deteriorates as the CSIs become completely outdated. Moreover, the results show that the impacts of the hardware impairments can be neglected at low SNR, however, the effects become more pronounced for high average SNR. Specifically, the performance analysis proves that TWTA is more severe compared to SSPA and SEL impairments models. The performance losses are measured in different ways such as the outage floor level, the capacity ceiling, and most importantly the severe reductions in diversity and multiplexing gains. Moreover, the results also show that the diversity gain can also experience some losses caused by the severity of the pointing errors, while it is not affected by the atmospheric pathloss. By comparing the rates achieved by mmWaves and sub-6 GHz for a big range of distance, we demonstrate that mmWaves are more suitable for densified small cells where the big data rate is highly required while sub-6 GHz is well relevant for big cells where the coverage requirement is of big interest. Furthermore, we studied the impacts of the interference density, the blockage density, and the NLOS pathloss exponent on the rate coverage, and we show that the performance is significantly vulnerable to the severety of these factors.  }

\appendices
\section{Proof of the achievable rate of the cellular system (\ref{rate})}
\textcolor{black}{After introducing the pdf expression of the effective SINR in (\ref{ratec1}), the achievable rate has the following generic integral form
\begin{equation}
    \mathcal{C}_1 = \int\limits_{0}^{\infty}x^{a-1}e^{-bx}\log(1+x)\text{d}x.
\end{equation}}

\textcolor{black}{The next step to transform the logarithm function into the Meijer-G function using the following identity \cite[Eq.~(07.34.03.0456.01)]{68}
\begin{equation}
\log(x+1) =  G_{2,2}^{1,2} \Bigg(x ~\bigg|~\begin{matrix}1,1 \\ 1,0 \end{matrix} \Bigg).  
\end{equation}}

\textcolor{black}{Applying the identity \cite[Eq.~(2.24.3.1)]{62} and after some mathematical manipulations, the achievable rate is finally derived as (\ref{rate})}.

\ifCLASSOPTIONcaptionsoff
  \newpage
\fi
\bibliographystyle{IEEEtran}
\bibliography{main}
\end{document}